\documentclass[a4paper,11pt]{article}
\pdfoutput=1 % if your are submitting a pdflatex (i.e. if you have
\usepackage{jcappub} % for details on the use of the package, please
\usepackage{amsmath,amssymb,mathtools,mathptmx}
\usepackage{longtable,multirow,multicol,tablefootnote}
\usepackage{longtable,multirow,multicol,tablefootnote}
\usepackage{txfonts,ae,aecompl}
\usepackage[normalem]{ulem}
\usepackage{graphicx,xcolor}
\usepackage[all]{hypcap}
\usepackage{appendix}
\usepackage{hyperref}
\usepackage{chngcntr}
\usepackage{orcidlink}
\definecolor{colo1}{RGB}{0,0,120}
\definecolor{colo2}{RGB}{0,100,80}
\definecolor{colo3}{RGB}{0,80,140}
\definecolor{colo4}{RGB}{90,90,90}

\hypersetup{colorlinks=true,
            linktoc=all,
            linkcolor=colo1,
            urlcolor=colo2,
            citecolor=colo3}

\newcommand{\hmpc}{{\, h^{-1}\, {\rm Mpc}}}
\newcommand{\hmpci}{{\, h\, {\rm Mpc^{-1}}}}

\newcommand{\cmpci}{{\, h^{3}\, {\rm Mpc^{-3}}}}
\newcommand{\kpc}{{\, {\rm kpc}}}
\newcommand{\Cn}{{\cal N}}

\newcommand{\Sk}{\cal S}
\newcommand{\Comb}[2]{{}^{#1}C_{#2}}
\newcommand{\expo}[2]{{#1} \times 10^{#2}}

\graphicspath{{images/}}

\title{Impact of Large-Scale Structure Formation on the Global Information Budget}

\author{Suman Sarkar\,\orcidlink{0000-0002-5465-3467}}

\affiliation{{\small Department of Physics, Indian Institute of Technology Kharagpur, West Bengal - 721302, India.}}

\emailAdd{suman2reach@gmail.com}

\abstract{This paper offers an original theoretical framework to quantify the information content associated with cosmological structure formation. By taking into account the growth in perturbations for the matter density field, we are able to calculate the relative reduction in the information entropy from the projection for a fully homogeneous setting. An analytical study is carried out to track back the evolution of global information content up to $z=20$, where a log-normal density distribution with redshift-dependent variance, skewness, and kurtosis is used to mimic the observable Universe. We report a $0.004\,\%$ drop in the information entropy of the Universe, that is caused by the formation of large-scale structures in the present universe. The accelerated expansion of the Universe is suspected as a countermeasure by nature to override the exponential growth in this entropic drop. We also analyse how the scale factor, expansion rate and growth rate of structure formation are connected with the entropic drop, taking the $\Lambda CDM$ model into account. The formalism is further developed and employed to study the spectrum of information underlying the galaxy distribution in the observable Universe. Using data from SDSS DR18 we also quantify the information sharing across different parts of the studied volume. An attempt to validate the assumption of cosmic homogeneity is made, which rules out the presence of a Universal scale of uniformity below $130 \hmpc$.}


\begin{document}
\maketitle
\flushbottom
\newpage
\section{Introduction}
%%%%%%%%%%%%%-----------------------------------------------------%%%%%%%%%%%%%%%%
The study of galaxy clustering is the cornerstone of modern cosmology. Over the years, it has played an important role in explaining the dynamics of the Universe. Peebles(1973) \cite{peebles73} introduced the primary statistical tools to characterise and quantify the properties of galaxy clustering. Later on, several seminal studies \cite{efstathiou79, hewett82, davis83, bardeen86, kaiser87, blanchard88, hamilton92, landy93, baugh93} refined our understanding of galaxy clustering and its manifestation across different length scales. Both observational facilities and theoretical frameworks have experienced substantial advancements in the last few decades. Large-scale surveys such as the SDSS \cite{york00}, 2MASS \cite{skrutskie06}, GAMA \cite{driver11} and DES \cite{abbott18} have not only expanded our observational capabilities in terms of individual object detection but also has contributed in understanding the current nature and evolution of galaxy clustering in the Universe. Furthermore, theoretical models and cosmological simulations such as Millennium Run \cite{springel05}, Horizon Run \cite{kim15}, EAGLE \cite{schaye15} and IllustrisTNG \cite{nelson18} have played a crucial role in interpreting observational data and uncovering the underlying physical mechanisms which govern galaxy clustering in the Universe.\\

On smaller scales, thermally supported interstellar clouds of self-gravitating gas fragments into stars \cite{jeans1914}. In the presence of anisotropic gravitational forces, a much larger, uniformly rotating or non-rotating cold gas cloud undergoes a spheroidal collapse to become a galaxy \cite{lynden62,lin65,binney77}. On the other hand, the spherical collapse model \cite{gunn72} explains how large galaxy clusters like \textit{Coma} can form through the infall of baryonic matter into dark matter (DM) potential wells. However, the formation of the filament and pancake-like structures, that are prevalent in the nearby observable universe, can be explained through the mechanism of anisotropic ellipsoidal collapse \cite{dorosh64,zeldo70,shandarin89}. Ostriker(1978) \cite{ostriker78} finds evidence of an inside-out dynamical growth of galaxy clustering, which is thought to be originating from the bottom-up hierarchical merging of DM halos \cite{press74,bond91,lacey93,sheth99}. Galaxies being the biased tracers of the DM halos are expected to be found in the sharpest density peaks of the DM distribution. To better understand the relationship between a galaxy, its associated DM host, and the space they occupy combined, it is necessary to comprehend the systematic evolution of the information content associated with the configuration of the large-scale structures occupying that space. Long before the galaxy's formation, this space was characterised by a nearly uniform matter distribution. Over time, gravity forces the matter to collapse and form a bound object, now seen as the galaxy. This process also involves the accretion of additional matter from relatively under-dense surroundings and indirectly helps the under-dense structures to grow. This entire process in every step reduces the randomness of the whole system ( galaxy + surrounding ) and takes it towards a low probability state. The birth of a galaxy requires several conditions to be met; only a few among numerous alternate possibilities would lead to the emergence of the galaxy within a specific space-time span. For example, a larger DM halo nearby could have changed the evolutionary trajectory of the galaxy in terms of space and time. The cosmic-web \cite{bond96} connects the galaxy clusters through filaments of galaxies that lie on the intersection of walls encompassing voids of various sizes \cite{gregory78,joeveer78,einasto80}. Hence, any movement of matter across space would cause perturbations in both the spatial distribution of probability and the continuum of information. \\

Building on the work of Nyquist(1924) \cite{nyquist24} and Hartley(1928) \cite{hartley28}, Shannon(1948) \cite{shannon48} put in place the groundwork for quantifying the information entropy in classical systems characterised by discrete variables. This led to subsequent research on modelling information exchange within complex systems \cite{jaynes57,wolfram83}. Information theory has long been used in astronomical data analysis. Using Fisher information Tegmark(1997) \cite{tegmark97a,tegmark97b} showed how to numerically estimate the accuracy at which observational data can determine cosmological parameters. Later on, Neyrinck(2009) \cite{neyrinck09} compared the information contents in the power spectra of the density and log-density fields. Concurrently, the information field theory \cite{ensslin09} was presented, reconstructing spatially distributed large-scale signals. Pandey(2013) \cite{pandey13} used Shannon entropy to develop the tool for testing the cosmological principle which was utilized in several studies to validate the cosmic homogeneity \cite{pandey15,pandey21} and isotropy \cite{sarkar19}. The Sufficient statistics to quantify the information content of a log-normal density distribution resulting from a Gaussian initial condition was developed in a contemporary work \cite{carron13}. Recently Pinho (2020) \cite{pinho20} quantified the information content in cosmological probes connecting information entropy to Bayesian inference. Information theory is currently being applied extensively in astronomical data analysis, targeting a lossless recovery of the cosmological information encoded with the incoming signals. Despite these efforts, a comprehensive information-theoretical technique to map the spatial information distribution of the Universe is still not in place. We develop a distinctive identity, the \textit{entropic gain} that quantifies the fraction of existing information in the Universe that goes into creating large-scale structures. Furthermore, the results presented in the paper aid in understanding the systematic change in the information content of the Universe throughout different redshifts. The evolution of Information entropy in inhomogeneous \cite{hosoya04} and homogeneous \cite{pandey19, das19, das23, pandey23} cosmological models has been extensively studied so far. In this work, however, we provide an entirely new methodology to analyze the entropic change of information resulting from the growth of perturbations in the matter density field, starting from a basic premise. In this study, we develop tools that not only allow us to measure the effective amount of spatial information but also help to determine the information sharing on various length scales associated with overall structure formation. This allows us to investigate the existence of a certain scale of homogeneity in the observable galaxy distribution. The cosmological principle (CP) that advocates the uniformity of the Universe on large scales, is a foundational assumption of modern cosmology that acts as a basis for the standard cosmological model. Several studies in the past have reported that the galaxy distribution in the observable Universe exhibits a fractal nature on small scales but gradually tends towards homogeneity beyond $50-90\hmpc $\cite{martinez94, borgani95, guzzo97, amendola99, hogg05, sylos07, dias23}. However, several studies suggest this fractal hierarchy \cite{pietronero87,coleman92,joyce99, sylos09, goncalves21} or shell-like distribution of galaxy groups\cite{einasto16} persist beyond $120 \hmpc$. Clowes et al (2013) \cite{clowes13} reports the existence of a Large Quasar Group (LQG) extending $> 500 \hmpc$; this questions the possibility of the cosmic homogeneity to lie anywhere around a hundred megaparsecs. However, subsequent studies \cite{nadathur13,sarkar16,sarkar16b} overrule this conclusion by reporting the scale of homogeneity in Luminous Red Galaxy (LRG) and quasar distributions to lie around $150-250 \hmpc$. Despite this, one cannot deny the presence of massive cosmic structures like the Sloan Great Wall\cite{einasto22}, Eridanus void\cite{kovacs22} or the Local hole\cite{haslbauer20}, in the observable Universe; this again raises questions on the assumption of cosmic homogeneity. As the opinions are still very much divided on the topic of cosmic uniformity, we try to shed light on this issue by using a novel estimator, namely the \textit{mutual gain}.\\

Throughout the paper, we adopt the $\Lambda CDM$ cosmology and use the following set of cosmological parameters \cite{planck18} for comoving distance calculation, N-body simulation and estimation of global information entropy: $\Omega_{m0} = 0.315$; \,\,  $\Omega_\Lambda = 0.685$; \,\,  $H_0 = 67.4$;  \,\, $\sigma_8 = 0.811$. \\

%%%%%%%%%%%%%%%%%%%%%%%%%%%%%%%%%%%%%%%%%%%%%%%%%%%%%%%%%%%%%%%%%%%%%%%%%%%%%%%%%%%%%%%%%%%%%%%%%%%%%%%%%%%%%%%%%%5
\section{Method of Analysis}
%%%%%%%%%%%%%-----------------------------------------------------%%%%%%%%%%%%%%%%

\subsection{Probability associated with perturbations in matter density field}
\label{sec:method_intro}
Let us start by considering $M$ objects distributed inside a cube of dimension $L \times L \times L$. We can divide the entire cube into $N_v$ voxels such that $N_v=N_s^3$. Here $N_s$ is the number of segments in each direction. So now, We have a mesh of $N_s^3$ grids with grid spacing $\Delta x =\frac{L}{N_s}$.\\

\noindent The $M$ objects inside the cube with $N_v$ voxels can be arranged in $\Omega$ different ways, when 
\begin{eqnarray}
\Omega &=& \frac{(N_v+M-1)!}{(N_v-1)!M!}.
\end{eqnarray}

\noindent Let us consider the case of finding $\mathrm{m}$ objects inside any of the $N_v$ voxels. Keeping $\mathrm{m}$ objects inside that voxel, the other $M-\mathrm{m}$ objects can be arranged across the $N_v-1$ voxels in $\omega_\mathrm{m}$ ways, given
\begin{eqnarray}
\omega_{\mathrm{m}} &=& \frac{\{(N_v-1)+(M-\mathrm{m})-1\}!}{(N_v-2)!(M-\mathrm{m})!}.
\end{eqnarray}

\noindent  The probability of finding $\mathrm{m}$ objects in a given voxel is then
\begin{eqnarray}
\label{eq:prob1}
p\,(\,\mathrm{m}\,)=\frac{\omega_\mathrm{m}}{\Omega}  \,= \, \frac{\{(N_v-1)+(M-\mathrm{m})-1\}! (N_v-1)!M!}{(N_v+M-1)(N_v-2)!(M-\mathrm{m})!},
\end{eqnarray}
with $\Cn=N_v+M-1$, \autoref{eq:prob1} becomes 
\begin{eqnarray}
\label{eq:prob2}
p\,(\,\mathrm{m}\,) &=& \left(\frac{N_v-1}{\Cn-\mathrm{m}}\right) \prod_{i=0}^{\mathrm{m}-1}{ \left(\frac{M-i}{\Cn-i}\right)}.
\end{eqnarray}
%-----------------------------------------------
\begin{figure*}
\centering
\includegraphics[width=\linewidth]{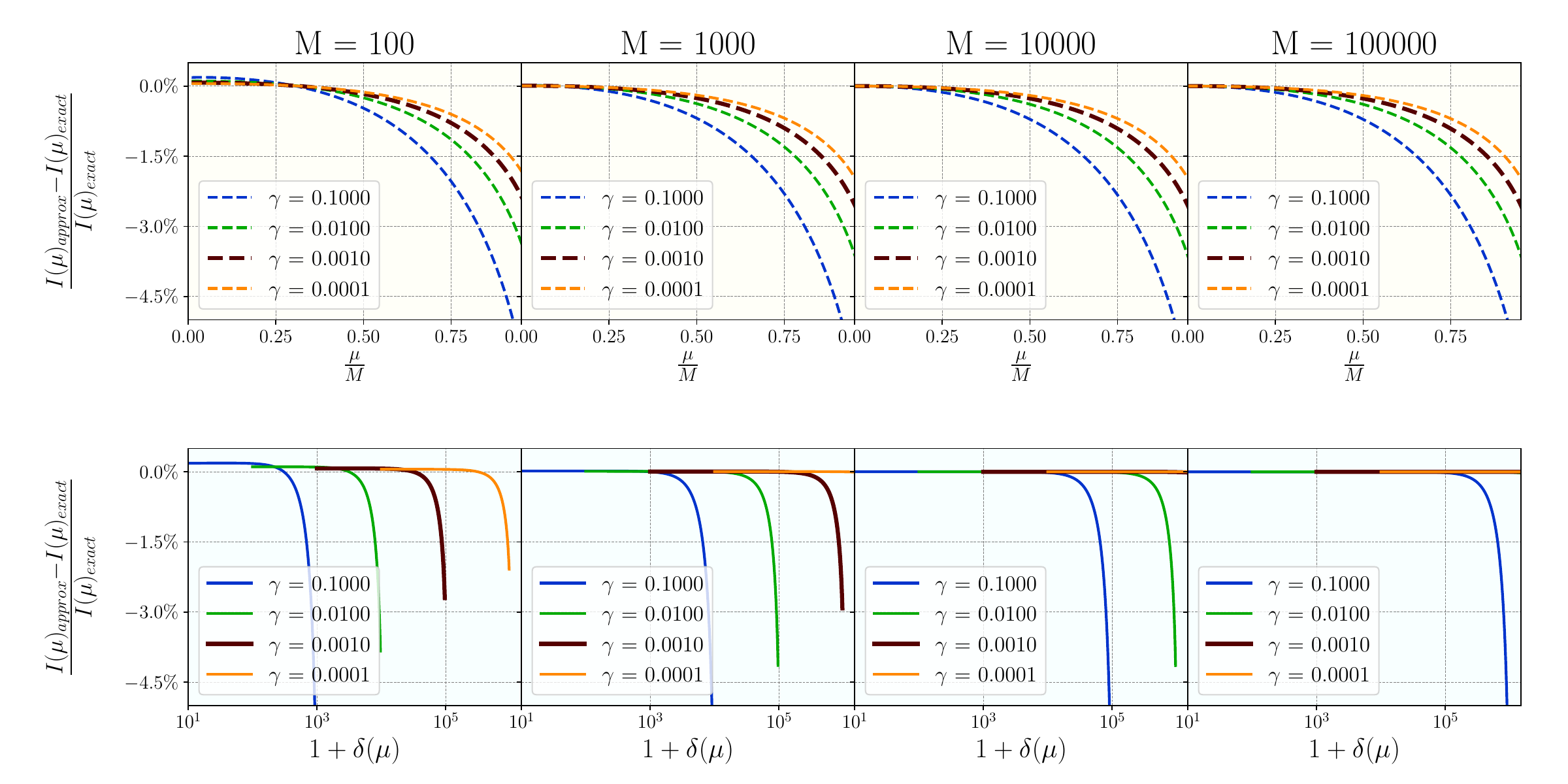}
\caption{This figure shows the effect of the approximation proposed in \autoref{eq:approx}}
\label{fig:approx}
\end{figure*}
%------------------------------------------------
\noindent This relation can be used to calculate the probabilities at each voxel in the mesh. However, with $\mathrm{m}$ as an integer, one can only deal with discrete distributions using \autoref{eq:prob2}. To evaluate the expected probabilities for any fractional real values of $\mathrm{m}$, one requires an approximation  
\begin{eqnarray}
\prod_{i=0}^{\mu-1}{ \left(\frac{M-i}{\Cn-i}\right)} & = & \left(\frac{M-\frac{\mu}{2}}{\Cn-\frac{\mu}{2}}\right)^\mu;
\label{eq:approx}
\end{eqnarray}
\noindent where $\mathrm{m}$ is replaced by $\mu$, a continuous real-valued variable within $0 \leq \mu < M$. The objective is to partition any continuous or discrete field into multiple virtual patches, referred to as \textit{``chunks"} hereafter, each possessing identical content but varying in morphology. Cumulatively, these chunks cover the entire volume in consideration. After employing the approximation, \autoref{eq:prob2} becomes 
\begin{eqnarray}
\label{eq:prob3}
p\,(\,\mu\,)  &=& \left(\frac{N_v-1}{\Cn-\mu}\right) \left(\frac{M-\frac{\mu}{2}}{\Cn-\frac{\mu}{2}}\right)^\mu.
\end{eqnarray}

\noindent Let us now introduce the parameters $\gamma$ and $\bar{f}_{\cal M}$ such that $\gamma=\frac{M}{N_v}$ is the \textit{mean chunk density} and $\bar{f}_{\cal M} = \,\langle \frac{\mu}{M} \rangle = \,\frac{1}{Ng}$ is the \textit{mean chunk fraction}. Also, $\mu$ can be written in terms of fractional density contrast $\delta$ as $\mu=(1+\delta)\gamma$. As there are ${\bar f}_{\cal M}^{-1}$ voxels in the entire cube, without prior knowledge of the density profile the probability of finding the density contrast $\delta$ for a given voxel is 
\begin{eqnarray}
\label{eq:prob4}
p\left(\,\delta\,\right) = \left[ (1+\gamma) - \frac{\gamma\delta\bar{f}_{\cal M} }{1-\bar{f}_{\cal M}} \right]^{\,-1} \,\, \left[ 1+ \frac{(1 -\bar{f}_{\cal M})}{ \gamma \left\{1-( 1+\delta)\frac{\bar{f}_{\cal M}}{2}\right\}}\right]^{\,-\,\gamma\,(\,1\,+\,\delta\,)}
\end{eqnarray}

%%%%%%%%%%%%%-----------------------------------------------------%%%%%%%%%%%%%%%%

\subsection{Information associated with density perturbations}
\label{sec:info_lss}
So far in this section, we have not mentioned galaxies or Large-scale Structures. We started with $M$ discrete point objects having unique locations in terms of voxel indices. However, using further approximations, we now have $M$ extended chunks that span throughout the volume in consideration. Hence, they can partially contribute to more than one voxel. Now, let us consider $G$ galaxies distributed in the cubic volume, comprised of $N_v$ grids. We can virtually divide $G$ galaxies into $M$ chunks of equal weightage, each having $ \frac{G}{M}$ galaxies; Hence, for any given voxel containing $g$ galaxies, the density contrast can be found as $\delta = \frac{g}{G\,\bar{f}_{\cal M}} -1 $.\\

\noindent \autoref{eq:prob4} exhibits the probability of finding the overdensity $\delta$. Now, probability is closely related to information. Probability quantifies the likelihood of an event and information measures the reduction of uncertainty in a system or process through the outcomes. The \textit{information content} corresponding to the probability of getting overdensity $\delta$ can be found as, 
\begin{eqnarray}
\label{eq:info4}
I\left(\,\delta \,\right) = \ln\left[  (1+\gamma) - \frac{ \gamma \delta \bar{f}_{\cal M}}{1-\bar{f}_{\cal M}} \right]+ \gamma(1+\delta) \ln \left[ 1+ \frac{(1 -\bar{f}_{\cal M})}{ \gamma \left\{1-( 1+\delta)\frac{\bar{f}_{\cal M}}{2}\right\}}\right].
\end{eqnarray}

\noindent  Here $I(\,\delta\,)$ is measured in the units of natural bits or \textit{nats}. \autoref{eq:info4} displays a crucial relationship between the density perturbation and the Information associated with the large-scale structures. Hereafter any mention of information would mean this large-scale information. \autoref{eq:info4} applies to both discrete and continuous fields since it no longer contains any explicit constraint of discreteness. When we started with $M$ and $N_v$ as integers, we could work only with discrete point data. Now, one can think of a continuous field that spans the entire volume but is virtually discretized into $M$ chunks of equal contribution. \autoref{eq:info4} allows to work with extensive variables that contribute fractionally to the elementary volume. Also, the shape of the entire region and the elementary volumes are no longer restricted to a specific geometry. However, one has to carefully choose ${\bar{f}_{\cal M}}$ and $\gamma$ depending on the clustering strength ($\delta$). The two terms in \autoref{eq:info4} on which the logarithm is being operated must produce non-zero positive numbers. This can be easily taken care of by choosing sufficiently large $N_s$ or very small values of ${\bar{f}_{\cal M}}$. We also have to remember that the approximation presented in \autoref{eq:approx} allows us to have the final form of \autoref{eq:info4}, and one has to be careful while using such an approximation. The limitation of this approximation is shown in \autoref{fig:approx}. For $\gamma<0.01$ and $M>100$ one can safely use the approximation with more than $99.9\%$ accuracy up to $\delta \leq 10^4$. However, for larger values of $\gamma$, it is necessary to have a large number of chunks in use.\\
%fffffffffffffffffffffff
\begin{figure*}
\centering
\includegraphics[width=\linewidth]{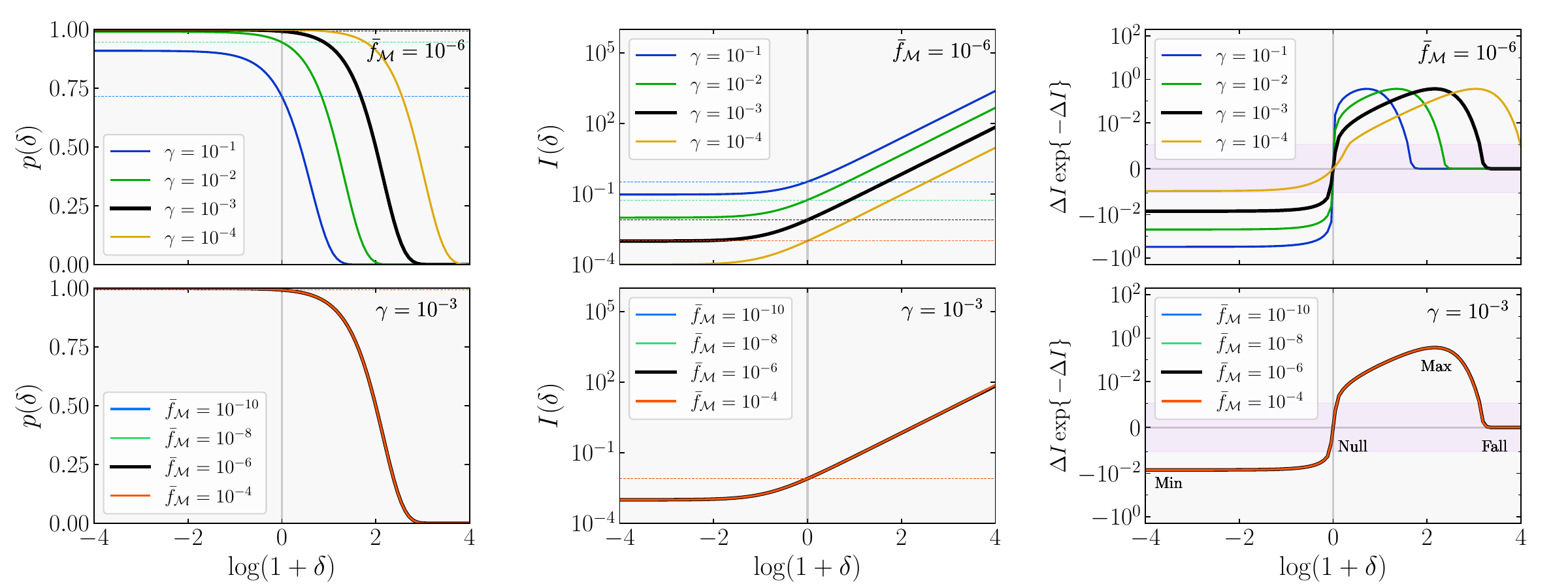}
\caption{This figure shows probability [left], information [middle] and entropic gain [right] as a function of $\delta$ for different values of $\gamma$  and $\bar{f}_{\cal M}$. The two panels on the right show the identity inside the square brackets in \autoref{eq:sp_info_gain}. Symmetric logarithmic scales are used along the vertical axis where the region with the purple shade has a linear scaling.}
\label{fig:prob_info}
\end{figure*}
%fffffffffffffffffffffff

\noindent For a completely homogenous distribution, we would have $\delta=0$; also for any voxel with a density same as the critical density of the Universe would have the information 
\begin{eqnarray}
\label{eq:Ibar1}
\bar{I} = \ln\left[\,1+\gamma\,\right] \,+ \,\gamma \ln \left[ 1+ \frac{1 -\bar{f}_{\cal M}}{\gamma \left(1- \frac{\bar{f}_{\cal M}}{2}\right)}\right]
\end{eqnarray}
where the associated probability is $\bar{p}= e^{-\bar{I}}$. The information offset from the homogeneous background can now be found as
\begin{eqnarray}
\label{eq:DI}
\Delta I(\,\delta\,) & = & I(\,\delta\,) \,-\, \bar{I}
\end{eqnarray}
The increment in information ($\Delta I$) at any given point originates from the relative change in the mass distribution of that point and the surrounding. It is reflected through the reduction in the probability ($\Delta p$) as $\Delta I =  - \ln \left[ 1+ \frac{\Delta p}{\bar{p}} \right] $. Hence, a -ve $\Delta p$ would result in a +ve $\Delta I$ and vice-versa.\\

It is imperative to consider that the mass distribution of the universe has been discretely characterized based solely on the presence or absence of galaxies. Since galaxies occupy a minuscule portion of the total space, it is more probable to encounter empty regions than substantial overdensities. When comparing a galaxy cluster to an individual void, owing to the abundance of voids in the Universe, a unit volume extracted from a void contains less information than its counterpart from a cluster. However, the collective contribution of information from voids is significant due to the higher likelihood. The departures from the mean background would lead to a change in information, which in the case of overdense structures goes into its internal configuration changing the degrees of freedom of the constituents. On the other side, the expanding space would offer more states for the chunks to be arranged, resulting in a growth of uncertainty and information. So, the extent and direction of the change in information entropy would vary between overdense and under-dense structures.\\

%%%%%%%%%%%%%-----------------------------------------------------%%%%%%%%%%%%%%%%
\subsection{Information entropy and large-scale structure formation}

\noindent At this point, we define the normalized \,\textit{Information Entropy} for the entire region as 
\begin{eqnarray}
\label{eq:info_ent}
{\cal E}^{I} &=&  \frac{1}{\bar{I}} \,\,\biggl[ \frac{ \int_V \,\,d^3 x \,\,\, p(\delta (\mathbf{x}) ) \,\,\, I(\delta (\mathbf{x}) )\,}{\,\,\int_V \,\, d^3x \,\,\, p (\delta (\mathbf{x} ))}\biggr].
\end{eqnarray}

\noindent However, the effective change in information entropy is measured through the \,\textit{Entropic gain}, defined as
\begin{eqnarray}
\label{eq:ent_gain}
{\cal G}^{I} & = & \frac{1}{\bar{I}} \,\,\biggl[ \frac{ \int_V \,\,d^3 x \,\,\, \Delta I(\delta ) \,\,\exp \{ -  \Delta I(\delta)\}\,}{\,\,\int_V \,\,d^3x \,\,\, \exp \{ -  \Delta I(\delta)\}}\biggr];
\end{eqnarray}
Note that $ {\cal E}^{I} = 1 + {\cal G}^{I}$, implying a reduction in total information entropy for a negative entropic gain. Assessment of the entropic gain does not require any presumption on the probability distribution of $\delta$. Despite having no prior knowledge regarding the profile of $\delta$, the collective excess information can be estimated by applying the appropriate weights to each elementary volume element based on the associated density contrast. \\

\noindent For a cubic region discretely divided into a finite number of grids with spacing $\Delta x$, one can measure The \textit{specific information gain} $\Phi \,(\,\mathbf{x}_i\, )$, i.e. the excess amount of information (weighted) per unit volume around $\mathbf{x}_i$. With $U_0 = \sum_{i} \,\exp \{\,-\Delta I\}$, the specific information gain can be expressed as
\begin{eqnarray}
\label{eq:sp_info_gain}
\Phi \,(\,\mathbf{x}_i\, )\, &=&  \frac{1}{\bar{I}\,U_0\,\Delta x^3} \biggl[\,\,\, \Delta I\,(\delta\, ( \mathbf{x}_i)\, ) \,\, \exp \{\,- \Delta I\,(\,\delta (\mathbf{x}_i)\,) \,\} \,\,\,\biggr].
\end{eqnarray}

\noindent For a completely homogeneous distribution, the specific information gain at any given point as well as the total entropic gain of the entire volume would be zero. We obtain +ve and -ve $\Phi$ for underdense and overdense structures respectively. The right panels of \autoref{fig:prob_info} show $\bar{I}\, U_0 \Delta x^3 \Phi$  or $\Delta I e^{-\Delta I}$ as a function of $\delta$ in symmetric log scale, whereas the left and middle panels show the probability and information associated to different overdensities. The Top panels show the dependence on density contrast when $\gamma$ is varied. The bottom panels show the dependence on $\bar{f}_{\cal M}$. The function $\Delta I e^{-\Delta I}$ has four characteristic points \textit{min}, \textit{null}, \textit{max} and \textit{fall} defining the curve, which are marked in the bottom right panel of \autoref{fig:prob_info}. The \textit{min}, \textit{null} and \textit{max} points are located at $\delta = -1$, $\delta = 0$, $\delta = \left[\gamma \ln \left( 1+ \frac{1}{\gamma}\right)\right]^{-1}$ respectively. From the bottom right panel, it is clear that $\Phi$ is not sensitive to the number of volume elements when $\gamma$ is fixed. To incorporate the effect of structure formation all over the Universe, one can keep $\gamma$ fixed and start adding voxels in all directions to extend the cube's volume to accommodate all the virtual chunks that span the Universe. This would require an enormously large number of voxels or an infinitesimal $\bar{f}_{\cal M}$. Imposing the condition ${\bar{f}_{\cal M}} \to 0$ leads to 
\begin{eqnarray}
\label{eq:I_u}
I (\,\delta\,) = \bar{I} + \delta \left[\gamma \ln \left( 1+ \frac{1}{\gamma}\right)\right].
\end{eqnarray}
With $\,\delta_c = \left[\gamma \ln \left( 1+ \frac{1}{\gamma}\right)\right]^{-1}$  and $\eta (\mathbf{x}) = \frac{\delta (\mathbf{x})}{\delta_c}$ \autoref{eq:ent_gain} becomes
\begin{eqnarray}
\label{eq:ent_gain_U}
{\cal G}^{U} & = & \frac{1}{\bar{I}} \,\,\biggl[ \frac{ \int_{{\mathbb{R}}^3} \,\,d^3 x \,\,\, \eta(\mathbf{x}) \,\,\exp \{ -  \eta(\mathbf{x})\}\,}{\,\,\int_{{\mathbb{R}}^3} \,\,d^3x \,\,\, \exp \{ -  \eta(\mathbf{x}) \}}\biggr];
\end{eqnarray}

\noindent Here ${\cal G}^U$ is the \textit{global information gain}. $\delta_c$ remains a free parameter unless it is constrained through experimental evidence. In this work, we choose $\gamma=\expo{8}{-4}$ to set $\delta_c \sim 175$ as this critical value represents the density threshold required for the virialization of a system. Past this threshold, the system configuration which has gradually developed over time through the externally acquired information, begins to be overridden by the internal stochasticity. Further increment in density contrast would cause the specific information gain to be suppressed and finally, it will diminish at the core ($\delta \to \infty$). \\

%%%%%%%%%%%%%-----------------------------------------------------%%%%%%%%%%%%%%%%

\subsection{Evolution of the global information entropy}
\label{sec:info_ent}
In this section, we conduct an analytical exercise to study the evolution of global information content starting from $z = 20$ up to $z = 0$ in the $\Lambda CDM$ Universe. We consider the growth of the perturbations to follow the form $\delta(\mathbf{x}, a)=D_{+}(a) \delta_0(\mathbf{x})$, as we focus on the linear perturbation regime by considering $\langle \delta^2_0(\mathbf{x}) \rangle = \sigma_8^2$. Here $\delta_0(\mathbf{x})$ is the spatial profile of density contrast at present and $D_{+}(a)$ denotes the growing mode of matter density perturbations, which can be expressed as 
\begin{eqnarray}
\label{eq:D_a}
D_{+} (\,a\,)= \frac{5\, \Omega_{m}(a)\,a^3\, \,E(a)^3}{2} \int^{a}_{0}{\frac{da^\prime}{{a^{\prime}}^3 E(a^{\prime})^3}},
\end{eqnarray}
where $\Omega_m(a)=\frac{\Omega_{m0}\,a^{-3}}{E(a)^2}$ is the matter density parameter at the scale factor $a$ with the normalised Hubble parameter as $E(a) = \frac{H(a)}{H_0} = \sqrt{ \Omega_{m0}\, a^{-3}+ (1-\Omega_{m0}) }$. \\

\noindent Before going forward, let us introduce the functional \textit{exponentially weighted moment} (EWM). The $k^{th}$ order EWM is defined as
\begin{eqnarray}
\label{eq:A_func}
{\cal A}_k [\,\eta\,]= \frac{\int_{\mathbb{R}^{3}} \, \{\, \eta(\mathbf{x}, a)\,\} ^k \,\exp \{\, -\eta(\mathbf{x}, a) \,\} d^3x }{\int_{\mathbb{R}^{3}} \, \exp \{\, -\eta(\mathbf{x}, a) \,\} d^3x }.
\end{eqnarray}

\noindent With $a$ as the scalefactor and $\eta(\mathbf{x}, a) = \frac{\delta(\mathbf{x}, a)}{\delta_c}$ as the scaled contrast, we can now find the global entropic gain (\autoref{eq:ent_gain_U}) in terms of the first EWM, i.e 
\begin{eqnarray}
\label{eq:entro_diff}
{\cal G}^{U} = \frac{{\cal A}_1 [\,\eta\,]}{\bar{I}}
\end{eqnarray}

\noindent Taking the derivative w.r.t the scale factor ( see \autoref{sec:appendix_1} ) leads to
\begin{eqnarray}
\label{eq:gain_diff}
\frac{d {\cal G}^U}{d a} &=& \frac{f(\Omega_m)}{a \,\bar{I}} \left[ {\cal A}_1^2 + {\cal A}_1- {\cal A}_2 \right],
\end{eqnarray}

\noindent where $f(\Omega_m)$ is the logarithmic growth rate $\frac{d \ln D_{+}}{d \ln a}$. \autoref{eq:gain_diff} can also be restructured to find the following identities
\begin{eqnarray}
\label{eq:gain_diff1}
\Gamma_{D+} = \frac{d \ln D_{+}}{d \ln {\cal G}^{U}} &=& \left[ 1 + {\cal A}_1 - \frac{{\cal A}_2}{{\cal A}_1} \right]^{-1}, \\
\label{eq:gain_diff2}
\Gamma_{a} = \frac{d \ln a} {d \ln {\cal G}^{U}}&=& \frac{1}{f(\Omega_m)} \left[ 1 + {\cal A}_1 - \frac{{\cal A}_2}{{\cal A}_1} \right]^{-1},\\
\label{eq:gain_diff3}
\Gamma_{H} = \frac{d \ln H}{d \ln {\cal G}^{U}} &=& - \frac{1+q}{f(\Omega_m)} \left[ 1 + {\cal A}_1 - \frac{{\cal A}_2}{{\cal A}_1} \right]^{-1}
\end{eqnarray}
Note that $q$ in \autoref{eq:gain_diff2} is the decelearation parameter, i.e. $q=-\frac{a\ddot{a}}{\dot{a}^2}$. At a given instant, the logarithmic slope $\Gamma_{X} = \frac{d \ln X} {d \ln {\cal G}^{U}}$ for an arbitrary variable $X$ relates the change in $X$ to the change in entropic gain through $\Delta X \,= \,\Gamma_{X}\,\Delta{\cal G}^U\,\left(\,\frac{X\, }{{\cal G}^U}\,\right)$. And for a constant $\Gamma_X$ we also have $X = X_0 \left( \,{\cal G}^U\,\right)^{\Gamma_X}$. Here the factor $\Gamma_X$ regulates the growth of $X$ at any instant, by controlling the sensitivity of $X$ on ${\cal G}^U$.\\

Let us now go back to \autoref{eq:A_func} and perform Taylor series expansion of the exponential terms, which gives the $k^{th}$ EWM in terms of the moments of $\eta(a)$; i.e.
\begin{eqnarray}
\label{eq:A_func_series}
{\cal A}_k = \frac{\sum_{i=0}^{\infty}{\frac{(\,-1\,)^i}{i!}}\,\langle \,\eta(a)^{i+k}\,\rangle}{\sum_{i=0}^{\infty}{\frac{(\,-1\,)^i}{i !}}\,\langle \,\eta(a)^{i}\,\rangle},
\end{eqnarray}
where 
\begin{eqnarray}
\label{eq:delta_hmom}
\langle \eta(a) \rangle = 0 \text{\,\,\,\,\,\,and\,\,\,\,\,\,}
\langle \eta(a)^2 \rangle = \frac{D_{+}^2(a)\,\,\sigma_8^2 }{\delta_c^2}.
\end{eqnarray}

\noindent We consider ($1+\delta$) to have a log-normal distribution with the skewness and kurtosis characterised by the $S_p$ parameters \cite{bernardeau95}. Hence, the $3^{rd}$ and $4^{th}$ order moments of $\eta$ will be 
\begin{eqnarray}
\label{eq:delta_hmom}
\langle \eta(a)^3 \rangle = \frac{S_3 \sigma^4}{\delta_c^3} \text{\,\,\,\,\,\,and\,\,\,\,\,\,}
\langle \eta(a)^4 \rangle = \frac{S_4  \sigma^6 + 3 \sigma^4}{\delta_c^4}.
\end{eqnarray}
Here $\sigma \,=\, D_{+} \sigma_8$ and the $S_p$ parameters at any given redshift $z$, are found as $S_3(z)=S_{3,0} (1+z)^{\alpha_3}$ and $S_4(z)=S_{4,0} (1+z)^{\alpha_4}$; where the present values of $S_3$ and $S_4$ are given as $S_{3,0} = 3 + \sigma_8^2$ and $S_{4,0}=16 + 15 \sigma_8^2 + 6 \sigma_8^4 + \sigma_8^6$ respectively. $\alpha_3$ and $\alpha_4$ control the rate of change of the skewness and kurtosis of the distribution at a given redshift. In this work, we have used $\alpha_3=-1.1$ and $\alpha_4=-2.3$ \cite{einasto21} to study the evolution of global information entropy from $z=20$ to $z=0$. \\

The quantities $\Gamma_{D_{+}}$, $\Gamma_{a}$ and $\Gamma_{H}$ presented in \autoref{eq:gain_diff1}, \autoref{eq:gain_diff2} and \autoref{eq:gain_diff3} in combination are useful for validating the $\Lambda CDM$ model and cosmological parameter estimation since ${\cal A}_1$ and ${\cal A}_2$ can be readily determined from observations. However, this work only focuses on the overall evolution of entropic gain for a log-normal density distribution through the $\Gamma$ parameters connecting the entropic gain with the expansion rate of the Universe and the growth rate of the LSS. \\

%%%%%%%%%%%%%-----------------------------------------------------%%%%%%%%%%%%%%%%
\subsection{Spatial distribution of Information}
\label{ssec:info_dist}
\begin{figure*}
\centering
\includegraphics[width=\linewidth]{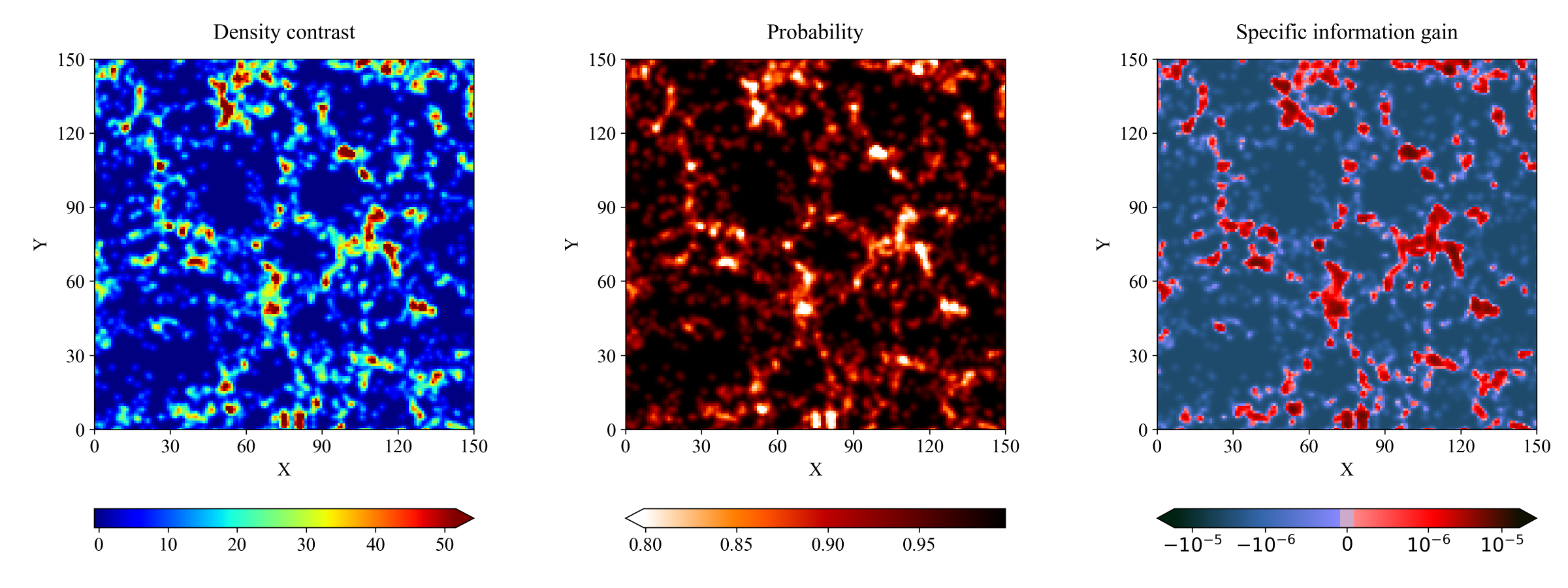}
\caption{The \textit{density}, \textit{probability} and \textit{specific information gain} is shown at different points of a $150 \hmpc$ cube from SDSS ( Sample 1 ). A $30 \hmpc$ slice in the $z$ direction is projected on the $X-Y$ plane.}
\label{fig:IG}
\end{figure*}

\noindent We conduct a numerical analysis to search the scale of homogeneity in the SDSS galaxy distribution. After a cube of size $L$ is carved out from the SDSS survey volume ( as described in \autoref{sec:data_sdss} ), it is divided into $N_s^{3}$ voxels of size $\Delta x$. The density contrast for each voxel is then found from the effective galaxy count. We estimate the \textit{specific information gain} for each voxel using the relation in \autoref{eq:sp_info_gain}. In this section, we do not impose $\bar{f}_{\cal M} \to 0$ and choose a finite value of $\bar{f}_{\cal M}$, as we are trying to validate the Cosmological principle itself. $\bar{f}_{\cal M}$ indirectly takes the cosmic homogeneity and isotropy into account, to assert that $\gamma$ remains constant when the boundary is extended to the horizon. that \autoref{fig:IG} shows the specific information gain at different locations within one of the cubes in Sample 1. It also presents how entropic gain correlates with galaxy clustering and respective probabilities. We use $N_s=50$ grids, setting $\bar{f}_{\cal M}=\expo{8}{-6}$. So, there are $100$ virtual chunks in use. \\
% %%--------------------------+++++++++++++++---------------------------%%%%
\subsubsection{Information spectrum} 
\label{sssec:info_spec}
\noindent After we have found $\Phi_{SNC}(\,\mathbf{x}_i\,)$, i.e. the shot noise corrected specific information gain ( see \autoref{sec:appendix_2} ) at each of the $N_v$ grids, we find the dimensionless 3D \textit{Information spectrum} as its Fourier transform
\begin{eqnarray}
\Sk (\mathbf{k}) &=& \int_{\mathbb{R}^{3}}  {\Phi_{SNC}( \mathbf{x} ) \, \exp \left\{ - 2 \pi i \,\mathbf{k} \cdot \mathbf{x} \right\}} \,\, d^3 x.
\end{eqnarray}

\noindent $\Sk$ is a dimensionless complex quantity that quantifies the amplitude of the entropic gain in different frequency ranges. It is numerically evaluated using the Discrete Fourier Transform (DFT), i.e.
\small
\begin{eqnarray}
    {\Sk}_{[l,m,n]} = \frac{1 }{\sqrt{N_v}} \sum_{l=0}^{N_s-1}\sum_{m=0}^{N_s-1}\sum_{n=0}^{N_s-1} \Phi _{[a, b, c]} \; \exp \left\{ \frac{-2 \pi i}{N_s} (a l  + b m  + c n ) \right\} 
\end{eqnarray}
\normalfont
\noindent Here $[a,b,c]$  are the indices for a specific voxel in the real-space; whereas, $[l,m,n]$ are indices for grids in Fourier space. The factor $\sqrt{N_v}$ ensures that Parseval's 
identity is satisfied. \\

\noindent The available wave numbers in the Fourier space for a grid spacing $\Delta x$ and box size $L= N_{s} \, \Delta x$ are   
\begin{eqnarray}
k_{m} & = & k_{N} \left[ \frac{2m}{N_s} - 1 \right]; \quad \quad  \text{for $\left\{ m = 0,1,2, \ldots,(N_s-1)\right\}$}
\end{eqnarray}
with $k_{N} = \frac{\pi}{\Delta x}$ as the Nyquist frequency.\\ 

\noindent Now, our goal is to find $\Sk$ as a function of $k$. We proceed by considering the entire available 3D $k$-space to be comprised of $N_b$ spherical shells centred at ($\frac{N_s}{2}$,$\frac{N_s}{2}$,$\frac{N_s}{2}$). Any $j^{th}$ shell is extended from $\frac{(j-1)\,\,k_{N} }{ N_{b}}$ to $\frac{j\,\,k_{N} }{ N_{b}}$, if $j$ takes values from $1$ to $N_b$. On the other hand, for any of the grid [$l$,$m$,$n$] in Fourier space, the magnitude of $k$ is found as $|\,k\,|= \sqrt{\, k_l^2+k_m^2+k_n^2 \,}$. The spherical bin that the grid belongs to is then $b =\lfloor \frac{N_{b} \, |\, k\, | }{k_{N}} \rfloor$. For each bin, we calculate the spherically averaged Information spectrum, $\bar{\Sk} (k) $ by taking a bin-wise average over $\Sk$, where each spherical bin has a specific value of $k$, associated with it. $\bar{\Sk}(k)$ now quantifies the average entropic gain across different frequencies.\\

%fffffffffffffffffffffff
%fffffffffffffffffffffff

% %%--------------------------+++++++++++++++---------------------------%%%%
\subsubsection{Information sharing across different length scales}
\label{sssec:info_share}
We perform an inverse-Fourier transform on $\bar{\Sk}$, normalizing it by the Nyquist frequency, to measure the entropic gain shared by the space on different length scales,
\begin{eqnarray}
{{\cal G}^{M}}(\,r\,) & = & \frac{1}{k_{N}} \int_{-\infty}^{+\infty} { W(k) \cdot \bar{\Sk} ( k ) \, \exp \left\{  2 \pi i k \cdot r\right\}} \,\, d k.
\label{eq:gr}
\end{eqnarray}
\noindent ${{\cal G}^{M}}(r)$ is called the \textit{Mutual gain}, and it quantifies the average amount of uncertainty reduced by structure formation on length scale $r$. $W(k)$ is the window function that mitigates the effects from the finite volume and geometry of the region. Throughout this analysis, We have used a \textit{Kaiser window} of the form 
\begin{eqnarray}
W(k) = \frac{i_0 \left(\,\, \beta \sqrt{1 - \left[\frac{k}{k_{N}}\right]^2}\,\,\right)}{i_0 \left( \,\,\beta \,\,\right) }.
\end{eqnarray}
Here, $i_0$ is the modified Bessel function of the zeroth-order and $\beta=14$. The simplified form of \autoref{eq:gr} in terms of DFT would be
\begin{eqnarray}
{{\cal G}^{M}}_{[a]} &=& \frac{1}{\sqrt{N_b}} \,\, \left[ \sum_{m=0}^{N_s-1}  W_{[m]} \, \, \bar{\Sk} _{[m]} \; \exp \left\{ \frac{2 \pi i \,a\, m}{N_b}  \right\} \right]
\end{eqnarray}
\noindent Mutual gain essentially tells how much of the extra information ( on average ) produced from the LSS formation is mutually shared by any two points in space. The magnitude of ${{\cal G}^{M}}(r)$ signifies the degree of uncertainty reduction attributed to one point in space, due to the exposure of the information stored at another point separated by a distance $r$. In the context of large-scale statistical homogeneity, by definition, the entropic gain diminishes with the absence of perturbations in the density field. Consequently, one would expect the mutual gain to exhibit a diminishing trend across the scale of homogeneity. 

%%%%%%%%%%%%%%%%%%%%%%%%%%%%%%%%%%%%%%%%%%%%%%%%%%%%%%%%%%%%%%%%%%%%%%%%%%%%%%%%%%%%%%%%%%%%%%%%%%%%%%%%%%%%%%%%%5

%%%%%%%%%%%%%-----------------------------------------------------%%%%%%%%%%%%%%%%
\section{Data}
%%%%%%%%%%%%%-----------------------------------------------------%%%%%%%%%%%%%%%%

\subsection{SDSS Data}
\label{sec:data_sdss}
\begin{figure*}
\centering
\includegraphics[width=\linewidth]{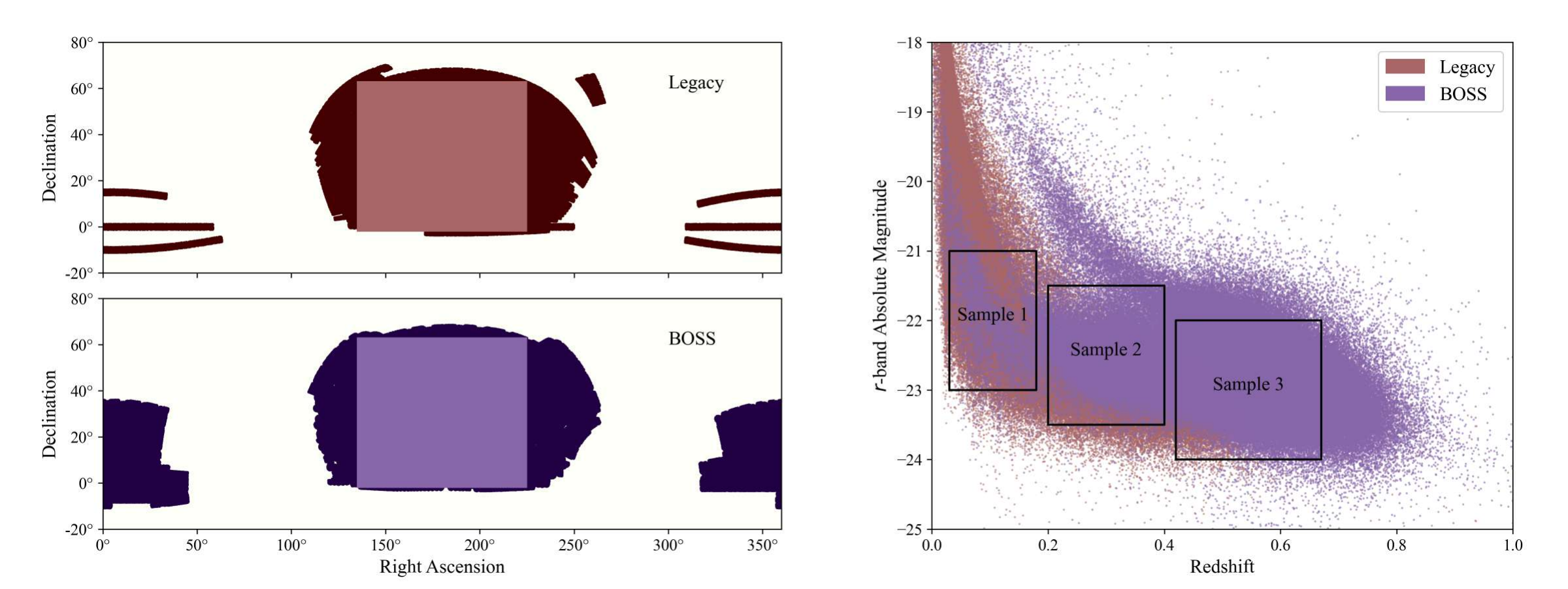}
\caption{[Left]: $RA$-$DEC$ span in the SDSS sky that is chosen for sample preparation, using BOSS and Legacy data. [Right]: Definition of the quasi-magnitude-limited samples on the $redshift$-$mangitude$ plain.}
\label{fig:SDSS_span}
\end{figure*}

We use data from the Sloan Digital Sky Survey (SDSS) to find the information content associated with galaxy clustering in the observable Universe. Galaxy samples from three 3 different redshift zones are prepared using data from the $18^{th}$ data release \cite{almeida23} of SDSS. DR18 incorporates refined data from all the prior programs of SDSS. However, we use data accumulated through the \textit{BOSS} and the \textit{Legacy} programs in this work. A \textit{Structured Query} is used to retrieve the required data from SDSS Casjobs \footnote{https://skyserver.sdss.org/CasJobs/}. To start with, conditions on the equatorial coordinates of the galaxies to choose galaxies within the angular span $135^{\circ} \le RA \le 225^{\circ}$ and $-2^{\circ} \le Dec \le 63^{\circ}$, as it provides a uniform sky coverage with high completeness. The left panel of \autoref{fig:SDSS_span} shows the sky coverage of both BOSS and Legacy; the rectangular region shows the chosen span. In the first step, we prepare 3 quasi-magnitude-limited samples by applying constraints over the redshift range and the (K-corrected + extinction-corrected) $r$-band absolute magnitude ( right panel of \autoref{fig:SDSS_span} ). The three conditions applied for the respective samples are tabulated in \autoref{tab:sdss_samples}. This minimizes the variation in number density in the samples, reducing the Malmquist bias, due to which fainter galaxies are difficult to identify at higher redshifts. 
%fffffffffffffffffffffff
%fffffffffffffffffffffff
%fffffffffffffffffffffff
\begin{figure*}
\centering
\includegraphics[width=\linewidth]{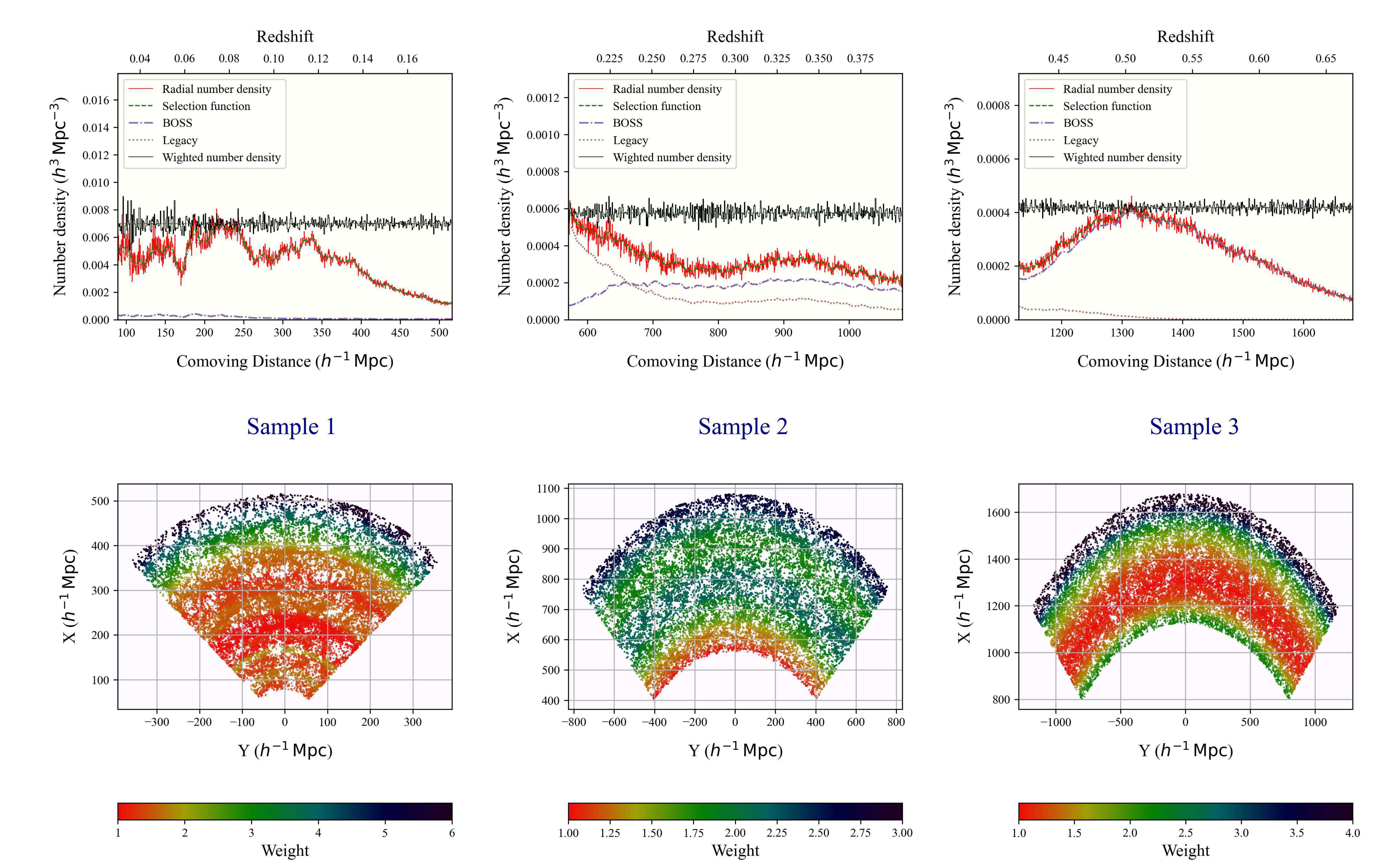}
\caption{[Top]: The number-density variation of the three samples before and after applying the selection functions. [Bottom]: The distribution of the samples projected on the $X-Y$ plane for slices ( in $Z$ direction ) of width 10 times the mean-intergalactic-separation. Point sizes in the display are proportional to the respective weights.}
\label{fig:sdss_num_vary}
\end{figure*}
%fffffffffffffffffffffff
%tttttttttttttttttttttt
\begin{table*}{}
\centering
\caption{Description of the three samples prepared from SDSS data}
\label{tab:sdss_samples}
\begin{tabular}{|l|l|l|l|}
\hline
&\textbf{ Sample 1} & \textbf{Sample 2} & \textbf{Sample 3} \\
\hline
\textbf{Absolute Magnitude} & $-23 \leq M_r \leq -21 $ & $ -23.5 \leq M_r \leq -21.5$ & $ -24  \leq M_r \leq -22$  \\
\textbf{Redshift range} & $0.03 \leq z \leq 0.18$ &  $0.2  \leq z \leq 0.4$  & $0.42  \leq z \leq 0.67$ \\
\textbf{Number density} & $\expo{6.87}{-3} \cmpci$ & $\expo{5.70}{-4} \cmpci$ & $\expo{4.13}{-4} \cmpci$ \\
\textbf{Size of cubes}& $150 \hmpc$ & $200 \hmpc$ & $250 \hmpc$ \\
\textbf{Stride} & $30 \hmpc$ & $40 \hmpc$ & $50 \hmpc$ \\
\textbf{Number of cubes} & $201$ & $1081$ & $922$\\ 
\textbf{Galaxies per cube} & $23074 \pm 2126$ & $4643 \pm 422$ & $6584 \pm 434$\\
\textbf{Avgerage galaxy size} & $13.19 \pm 4.66 \kpc$ & $23.95 \pm 7.42 \kpc$ & $24.05 \pm 10.45 \kpc$\\
\textbf{$u-r$ color} & $2.43 \pm 0.72$ & $3.78 \pm 1.90$ & $2.94 \pm 2.03$\\
\textbf{Redshift} & $0.116 \pm 0.021$ & $0.306 \pm 0.033$ & $0.542 \pm 0.039$\\
\hline
\end{tabular}
\end{table*}
%tttttttttttttttttttttt
Note that the BOSS data we use here includes targets from both LOWZ ($0.15 < z < 0.43$) and CMASS ($0.43 < z < 0.70$). On the other end, we have Legacy data with higher incompleteness on small redshifts ($z<0.02$). Also, fibre collision will affect the selection of galaxies in close vicinity. Hence, choosing an appropriate magnitude and redshift range alone does not provide a completely uniform sample selection. The radial number density of the combined (BOSS+Legacy) distribution still exhibits background fluctuations in addition to the intrinsic variation in number density ( resulting solely from the growth of inhomogeneities in the matter density field ). To decouple the two effects we identify the radial selection function $n(r)$ by applying a Gaussian kernel ( $5\sigma$). The smoothed radial number density field now considered the selection function, represents the residual Malmquist bias that has to be filtered out. Using this selection function we assign a weight $w(r)$ to each galaxy lying in a given radial bin $r$ to $r+\delta r$. If $\bar{n}_g$ is the maximum value of the $n(r)$ within the entire range of radial comoving distance then $w(r) = \frac{\bar{n}_g }{ n(r)}$. One must take the sum over the weights instead of adding a unit count for each galaxy. Put differently, for each galaxy found at the comoving radius $r$, we are also losing $w(r)-1$ neighbouring galaxies due to the limitations of the survey. The top panels of \autoref{fig:sdss_num_vary} show the radial number density profile of the three samples, before and after applying the selection function. The bottom three panels show galaxy distributions in slices, from each of the three quasi-magnitude-limited samples, projected on the $X$-$Y$ plane; each point represents a galaxy and the point sizes are scaled proportionally with the respective weights. Note that the three samples have different values of $\bar{n}_g$; hence, the values of the weights are specific to the samples. The comoving Cartesian coordinates of each galaxy in the sample are found using the spectroscopic redshifts and equatorial coordinates. After we have the three samples, we carve out cubic regions of $150 \hmpc$, $200 \hmpc$ and $250 \hmpc$ from \textit{Sample 1}, \textit{Sample 2} and \textit{Sample 3}, respectively. we curve out as many cubes as possible by taking successive strides in all three directions. Any two adjacent cubes share $80\%$ the side length or almost $50 \%$ of their volume. Relevant information about three samples is tabulated in \autoref{tab:sdss_samples}. For each sample, we have shown the (average) apparent size and $u-r$ color of the constituent galaxies. The size is determined from the product of the angular diameter distance and the Petrosian radius ( in radians ) of the circle containing $90\%$ of its total light.

% %%%%%%%%%%%%%-----------------------------------------------------%%%%%%%%%%%%%%%%
\subsection{Nbody Data}
\label{sec:nbody}
We generate multiple realisations of dark matter distributions by running a cosmological N-body simulation with GADGET4 \cite{springel21}. The simulation is performed by populating $128^3$ dark-matter particles inside a $500 \hmpc$ cube and tracking their evolution up to redshift zero. Ten realisations are generated and eight disjoint cubes are further extracted from each by dissecting them in each direction. We finally have $80$ cubes of size $250 \hmpc$ for each of the redshifts that we have access to. To analyse the evolution of information content for $\Lambda CDM$ cosmology we analyse the distributions of dark-matter particles at redshifts $0$, $0.5$, $1$, and $2$. For each of the $80$ cubes, we randomly select $100000$ dark-matter particles to retain them in the final samples. 
%fffffffffffffffffffffff
\begin{figure*}
\centering
\includegraphics[width=\linewidth]{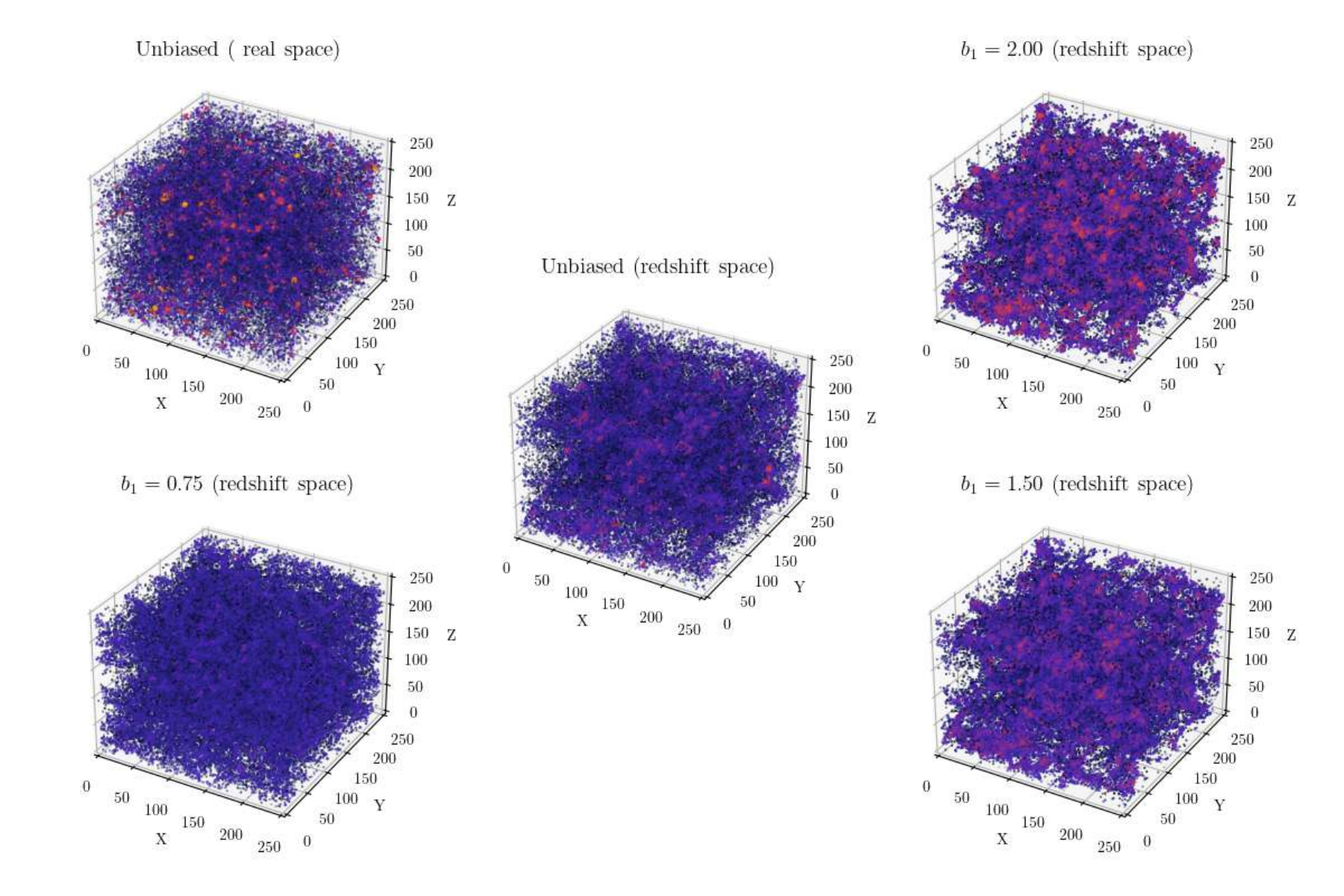}
\caption{The distribution of the different mock samples ($z=0.5$) described in \autoref{sec:bias_rsd}, before and after applying RSD and bias. Distributions for three different bias values are shown along with the original distribution and its unbiased counterpart in redshifted space. The variation in density contrast is represented using the variable colour palette.}
\label{fig:baised_dist}
\end{figure*}
%fffffffffffffffffffffff
% %%%%%%%%%%%%%-----------------------------------------------------%%%%%%%%%%%%%%%%
\subsection{Biased distributions in redshift-space}
\label{sec:bias_rsd}
% %%--------------------------+++++++++++++++---------------------------%%%%
\subsubsection{redshift-space mapping}
We take each of the $80$ N-body cubes for $z=0.5$ and introduce linear redshift-space distortions (RSD) by randomly placing the cubes within the accessible radial and angular span available for the SDSS data. The cubes are placed in such a way that the relative distance and orientation between the observer and the N-body cube centres mimic that of the SDSS (Sample 3) cubes. Using the velocities of each of the particles in the distribution we find their redshift-space coordinates as
\begin{eqnarray}
    \mathbf{s} = \mathbf{r} \, \left[ 1 + \frac{\mathbf{v} \cdot \mathbf{r}}{a\,H(a) \,\,||\mathbf{r}||^{2} } \right],
\end{eqnarray}
where $\mathbf{v}$ and $\mathbf{r}$ are the actual velocity and position vectors in real-space. $z=0.542$ is used for finding the scalefactor ($a$) and Hubble parameter, $H(a)$ to plug into the above equation.\\
%%--------------------------+++++++++++++++---------------------------%%%%
\subsubsection{Biased distribution}
\noindent Galaxies are formed after baryonic matter follows the pre-existing DM halos to select the density peaks for consolidation preferentially. So they are considered to be positively biased by and large. We use the N-body data to mimic a couple of biased galaxy distributions along with unbiased and negatively biased ones as well. To apply the bias on the redshift-space distribution of DM particles, we use the selection function (model 1) proposed by \cite{cole98}. For a given cube of dark-matter particles mapped in redshift-space, We apply a Cloud-In-Cell (CIC) technique with a mesh of $N_c^3$ grids to measure the density contrast ($\delta$) on the grids. Thereafter, the  z-score of density contrast, $\nu(x)=\frac{\delta(x) - \mu}{ \sigma} $ is found for each of the particles, where $\sigma = \sqrt{ < \delta ^2 >}$ and $\mu=\langle \delta \rangle = 0$. We then apply trilinear interpolation to determine $\nu$ at the particle positions from the $8$ adjacent grids. Next, the selection probability 
\[
p_s(\nu) \, = \, 
\begin{cases}
    \exp \,(\,A\nu + B\nu^{\frac{3}{2}}\,),     & \text{if } \nu \ge 0 \\
    \exp\,(\,A\nu\,) ,          & \text{if } \nu \leq 0  
\end{cases}
\]
%tttttttttttttttttttttt
\begin{table*}
\centering
\caption{Average bias for the $80$ mock biased distributions obtained from the redshifted N-body cubes.}
\label{tab:bias_tab}
\begin{tabular}{|c|c|c|c|}
\hline
\multirow{2}{*}[0.5em]{$A$} & \multirow{2}{*}[0.5em]{$B$} & \multicolumn{2}{c|}{Linear bias ($b_1$)} \\
\cline{3-4}
& & Target & Model \\
\hline
$1.0$ & $-2.51$ & $0.75$ & $0.755 \pm 0.008$\\ 
$1.5$ & $-1.11$ & $1.50$ & $1.505 \pm 0.014$\\ 
$2.0$ & $-1.20$ & $2.00$ & $1.990 \pm 0.019$ \\ 
\hline
\end{tabular}
\end{table*}
%tttttttttttttttttttttt
is applied to have a density-wise biased selection of the particles from the entire cube. We also find the base probability of the particles $p_0(\nu)=\frac{(1+\nu \sigma)}{N_c^3}$; i.e. the probability of finding a randomly selected particle to have contrast $\nu$. After applying the selection function, the selection probability of a particle with scaled contrasts $\nu$ into the final set is $p_f(\nu) = p_0 (\nu) \, p_s(\nu)$. Next, a random probability $p_r$ between 0 and $max\{p_f(\nu)\}$ is generated for each particle in the distribution. The particles which satisfy the criteria $p_r < p_f$ are selected to represent the galactic mass. We randomly select $100000$ particles, to get the final samples from each biased and unbiased distribution for further testing effects of linear RSD and biasing. The density contrast ($\delta$) is calculated ( by using CIC ), for each particle in the newly formed biased samples. Finally, linear bias is estimated by comparing their $\delta$ profile with the unbiased sample. We use different combinations of $A$ and $B$ and then check the linear bias for the distribution using the technique described in \ref{sec:appendix_3}. The parameters are tweaked to achieve the desired values of $b_1$ through trial and error. The suitable values of the parameters $A$ and $B$ which provide the desired values of linear bias $b_1$ are enlisted in \autoref{tab:bias_tab}. \autoref{fig:baised_dist} helps compare the redshift-space biased and unbiased distributions with the original real-space N-body data. All distributions have the same number of particles.  
%%%%%%%%%%%%%%%%%%%%%%%%%%%%%%%%%%%%%%%%%%%%%%%%%%%%%%%%%%%%%%%%%%%%%%%%%%%%%%%%%%%%%%%%%%%%%%%%%%%%%%%%%%%%%%%%%%
\section{Results}
%%%%%%%%%%%%%-----------------------------------------------------%%%%%%%%%%%%%%%%
\subsection{Evolution of Global Information entropy}
\begin{figure*}
\centering
\includegraphics[width=\linewidth]{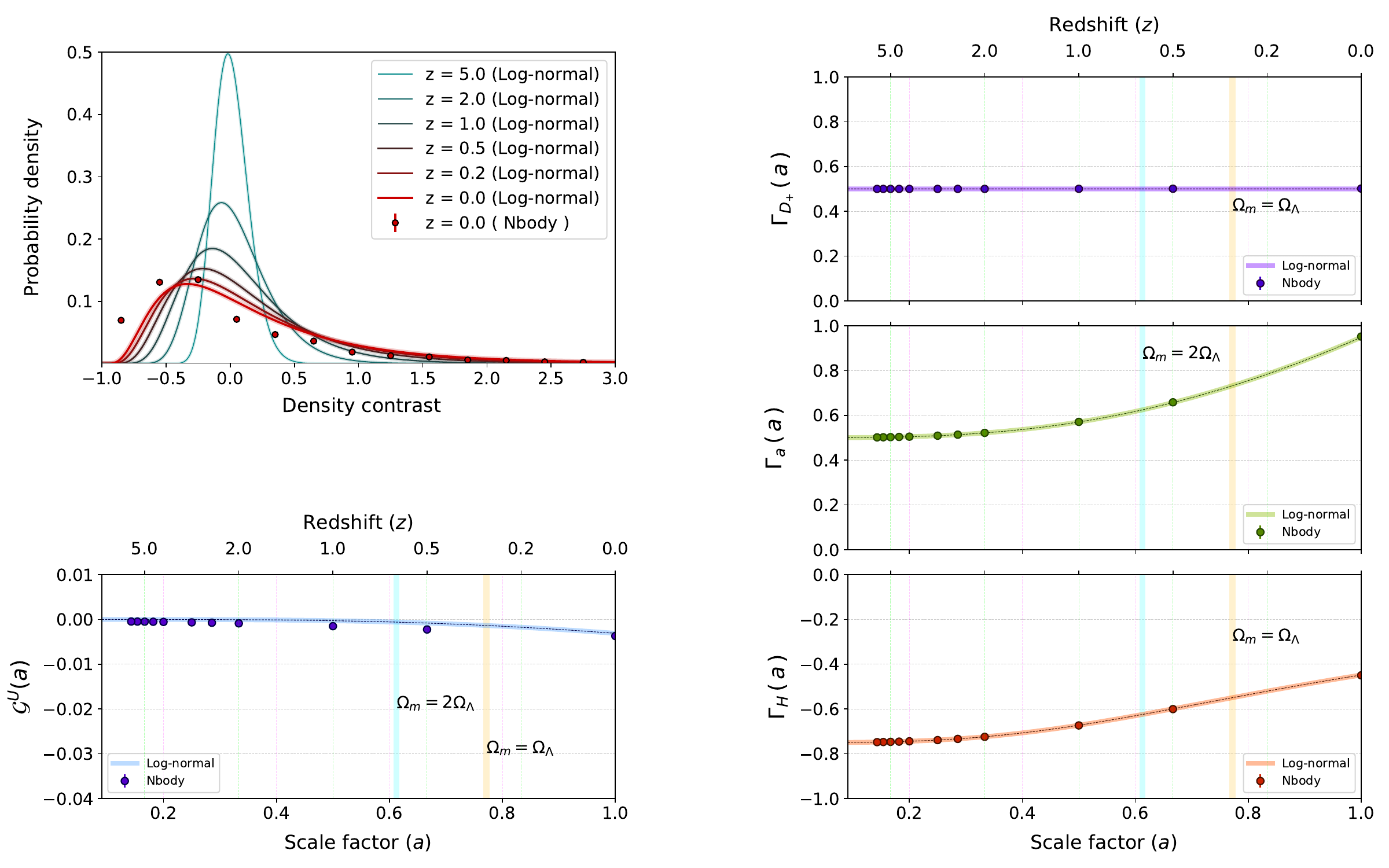}
\caption{[Top-left]: Evolution of the density field. [Bottom-left]: Entropic gain as a function of scalefactor. [Top-right]: $\Gamma_{D_{+}}$ as a function of scalefactor. [Right-middle]: $\Gamma_{a}$ as a function of scalefactor. [Bottom-right]: $\Gamma_{H}$ as a function of scalefactor.}
\label{fig:evolution}
\end{figure*}

In \autoref{sec:info_ent} we discuss the method to measure the global information entropy at a given epoch for a log-normal density distribution and introduce three dimensionless quantities $\Gamma_{D_{+}}$, $\Gamma_{a}$ and $\Gamma_{H}$ to study the dynamics of the matter and space of the universe through the entropic gain. In the right-hand side of \autoref{fig:evolution} we show these three quantities as a function of redshift or scalefactor. In the top left panel of \autoref{fig:evolution} we show the evolution of the density distribution responsible for causing the increment in the global entropic gain. The density distribution estimated for the $250 \hmpc$ ( with voxel size $\sim 8 \hmpc$ ) cube for the Nbody dark matter distribution is shown alongside the log-normal distribution ($z = 0$) for comparison. In the bottom left panel of  \autoref{fig:evolution} we show the global entropic gain as a function of the scale factor. A negative entropic gain or \textit{entropic drop} indicates the dominance of underdense regions in the governance of the large-scale configuration of the Universe. The nonlinear growth in the entropic drop illustrates the increasing growth of inhomogeneities in the density field. However, the net effect of entropic gain up to $z=0$ is minuscule, i.e. $ < 1 \%$ of the total information entropy of the Universe. On the top-right panel, we show $\Gamma_{D_{+}}$  as a function of scalefactor. We find that throughout all redshifts the growth rate of the Universe scales as $\sim \sqrt{{\cal G}^U}$. On the right-middle panel of \autoref{fig:evolution} we show the evolution of $\Gamma_{a}$. We find that the size of the Universe scales as $\sim \sqrt{{\cal G}^U}$ on large redshifts. The dependence of the expansion of space on the entropic drop gradually increases with time and on smaller redshifts the size of the Universe approaches a scaling of $\sim {\cal G}^U$. In the bottom-right panel, we show the variation of $\Gamma_{H}$ with scale factor. We note that the comoving expansion rate reduces with the increment in the entropic drop. An implication of this can be the increasing expansion rate which may result from the natural tendency of the cosmos to try and override the entropic drop. On larger redshifts, we get $\Gamma_{H} \sim {-0.75}$. As the Universe transitions from decelerated to accelerated expansion, $\Gamma_{H}$ has the maximum rate of change; it reduces again on smaller redshifts. The Hubble flow is comparatively less sensitive to the entropic drop at present with $\Gamma_{H} \sim {-0.45}$. Overall, the evolution of $\Gamma_{D_{+}}$, $\Gamma_{a}$ and $\Gamma_{H}$ for the N-body distribution is found to be almost identical to the log-normal distribution, which suggests that the change in these quantities are solely dependent on the choice of the cosmological model.

% fffffffffffffffffff
% fffffffffffffffffffffff
%%%%%%%%%%%%%-----------------------------------------------------%%%%%%%%%%%%%%%%

\subsection{Information spectrum and Information Sharing}
In \autoref{ssec:info_dist}, we discuss how to track the distribution of specific information gain in different frequencies through the information spectrum. On the other hand, we present the mutual gain to measure the average information gain shared between two points separated by a certain distance in space. At first, we take the 3 samples from SDSS and analyze the distribution of information in the observable Universe. We carry out the same analysis for simulated N-body data at different redshifts and further check the effect of linear RSD and bias by comparing SDSS results with redshift-space biased N-body mock distributions. For each case, we search for the scale of homogeneity where the mutual gain goes to zero. 

\subsubsection{Information spectrum in SDSS}
\begin{figure*}
\centering
\includegraphics[width=\linewidth]{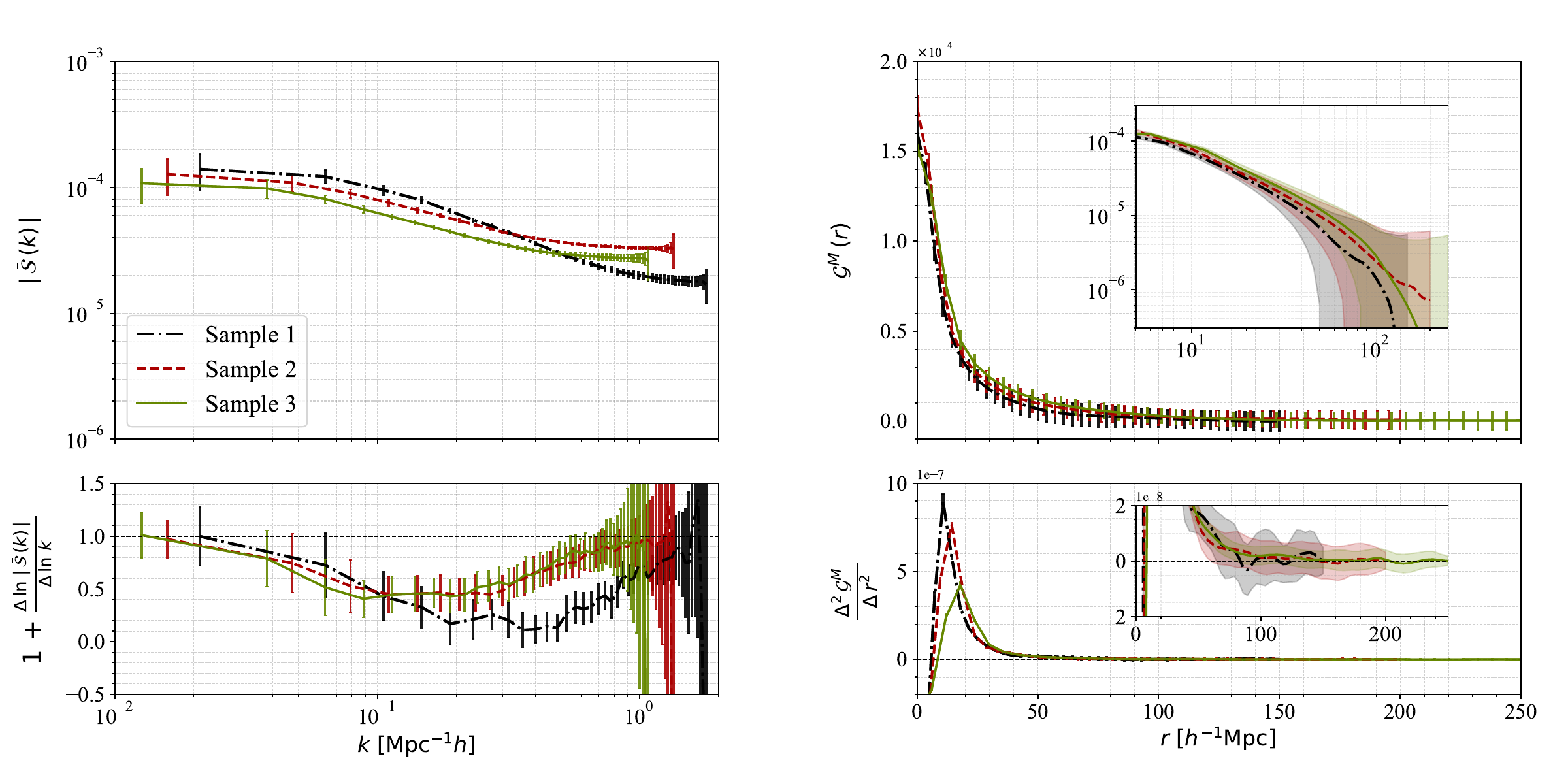}
\caption{[Top-left]: Information spectrum as a function of wave numbers.  [Bottom-left]: This figure shows the spectral index $n_s$, at different values of $k$. [Top-right]: Mutual gain as a function of length scales. The same thing is shown in the inset but in the log scale. [Bottom-right]: $2^{nd}$ derivative of mutual gain with length scale; a zoomed-in view is presented in the inset.}
\label{fig:Ik_Ix_sdss}
\end{figure*}

In the top-left panel of \autoref{fig:Ik_Ix_sdss}, the spherically averaged \textit{information spectrum} is presented for three samples representative of three redshift zones. We notice a higher degree of information gain is associated with the larger wavelengths (small $k$) for all three samples. The amplitude gradually falls with increasing $k$ and finally reaches a plateau at the high-frequency end. We notice an abrupt change in the slope of the information spectrum near $0.05 \hmpci$ for all three samples. This fall in spectral amplitude persists up to the nonlinear regime ($ > 0.8 \hmpci$); after which, the consistent drop in information slows down and remains steady up to the smallest scales probed. This non-diminishing tail is the characteristic signature of the information spectrum. Due to the presence of the exponential term, it contains all higher-order moments of density perturbations. Hence, it is capable of tracing the collective information offsets caused by all the multi-point correlations existing in that space of relevance. However, it is not capable of extracting the information on phase differences between the different modes. In \autoref{fig:Ik_Ix_sdss} (top-left panel) we find an overall trend of the information spectrum decreasing with increasing redshift. Sample 2 has a nearly identical profile as Sample 3 but a higher amplitude throughout. Similarly, Sample 1 has a higher amplitude than both Sample 2 and 3 up to $k=0.2 \hmpci$. Beyond $k=0.2 \hmpci$ we notice a rapid fall of the spectral amplitude for Sample 1 which is non-intuitive at first sight. On lower redshifts, the amplitude of information gain is expected to get higher with the enhancement in the definition of the structures. However, for Sample 1, we observe the information gain to be falling more rapidly in the large $k$ range. The presence of redshift-space distortions partly explains this fall, which is discussed later on in detail. That said, Sample 1 is a collection of galaxies from a different class than the other two samples, which could be a significant contributing factor to this drop in information gain on higher frequencies. From \autoref{tab:sdss_samples}, we find that galaxies in Sample 1 are much smaller in size compared to the galaxies in the other two samples. A relatively higher absolute magnitude and bluer $u-r$ color also indicate the presence of a younger population in the galaxy sample. This means that, unlike the other two samples, Sample 1 consists of a fair number of field and satellite galaxies that are not representative of a higher degree of clustering and are more uniformly distributed on small scales, compared to galaxies in Sample 2 and 3. \\

The bottom-left panel of \autoref{fig:Ik_Ix_sdss} shows the spectral index of information gain. If the information spectrum follows the relation $\bar{\cal S} \propto k^{n_s-1}$, then the spectral index of information gain is given as $n_s (k) = \frac{\Delta \ln {\bar{\cal S}}}{ \Delta \ln k}+1$. For all the samples, we find $n_s$ lying between $0$ and $1$, with a \textit{red-tilt} that points toward an inside-out picture of information sharing. The spectral index starts with $n_s=1$ at the smallest $k$ we probe. This shows a scale-independent nature of the information spectrum on the largest scales or earliest epochs. It then gradually starts reducing, eventually reaching a turnaround point where $n_s$ finds its minimum value. This is where the information gain has the maximum scale dependence. For Sample 2 and Sample 3, we can find this turnaround near $ 0.2 \hmpci$. For Sample 1 however, we see this drop with a much lower value of the spectral index ($n_s \sim 0.1$) and in a relatively higher frequency range ($ 0.3 -0.5 \hmpci$) where $\bar{\cal S}$ becomes almost inversely proportional to $k$. In other words, \textit{bigger the structure, greater the information it contains}. \\

%fffffffffffffffffffffff
%fffffffffffffffffffffff
%fffffffffffffffffffffff
\begin{figure*}
\centering
\includegraphics[width=\linewidth]{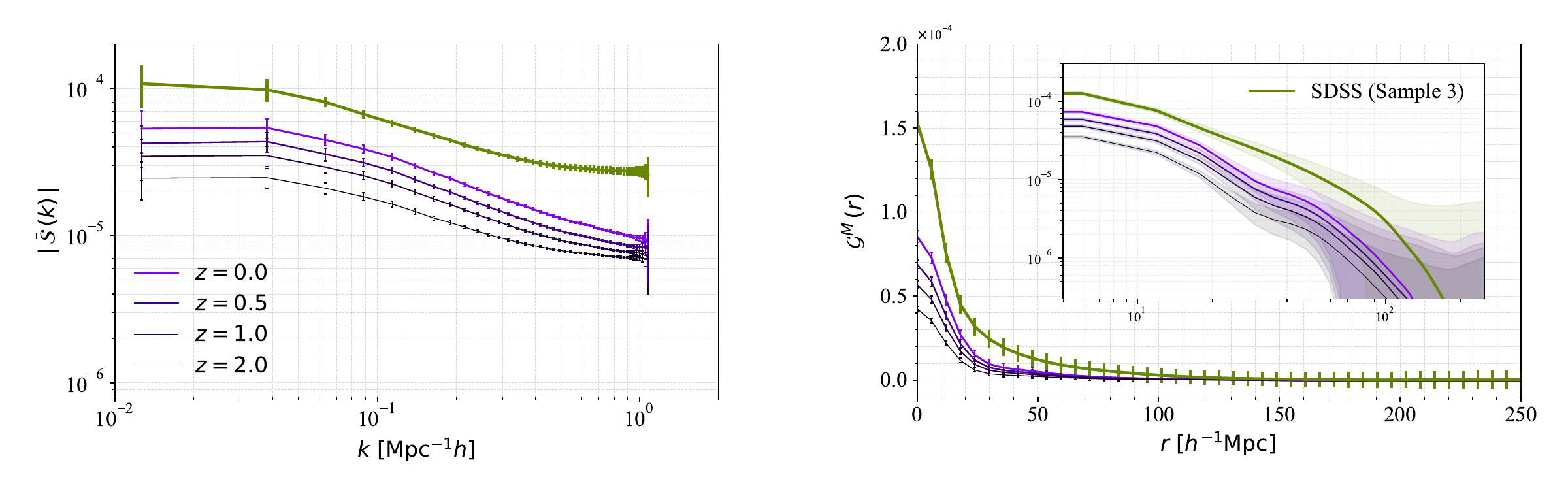}
\caption{[Left]: Information spectrum as a function of wave numbers for the N-body dark-matter distributions simulated at different redshifts. The SDSS data (Sample 3) is shown alongside. [Right]: Information sharing as a function of length scales, for the same distributions.}
\label{fig:Ik_Ix_nbody}
\end{figure*}
%fffffffffffffffffffffff
%fffffffffffffffffffffff
\begin{figure*}
\centering
\includegraphics[width=\linewidth]{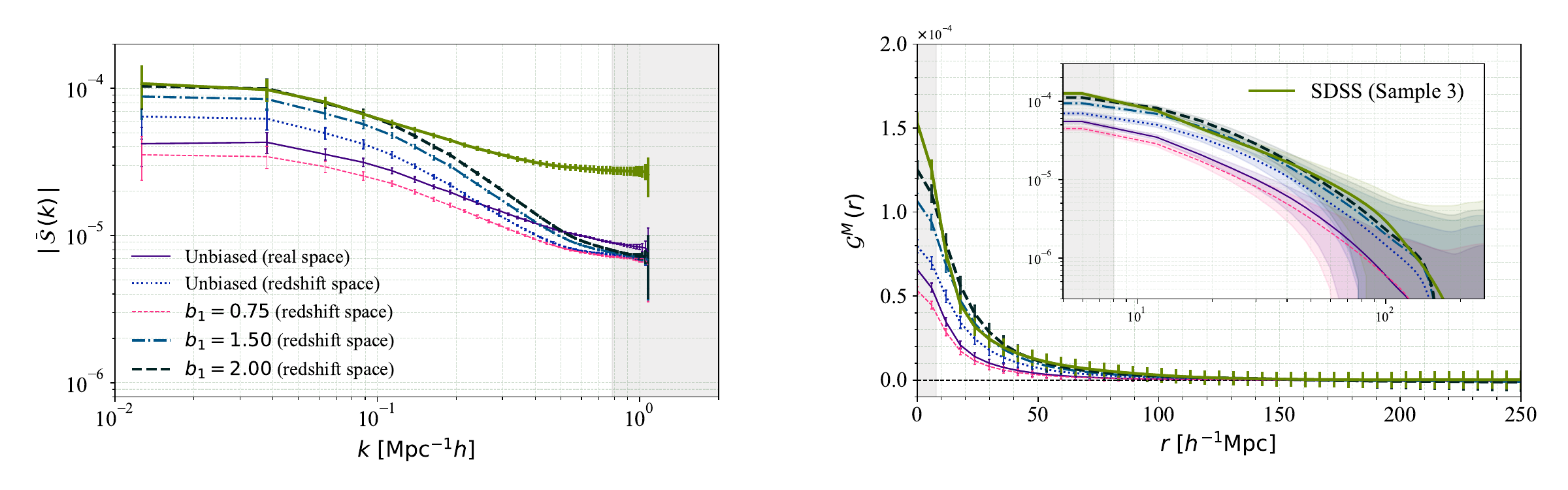}
\caption{Same as \autoref{fig:Ik_Ix_nbody} but for redshift-space biased distributions. The original real-space distribution, the redshift-space unbiased distribution and the SDSS Sample 3 are also shown for comparison.}
\label{fig:Ik_Ix_rsd}
\end{figure*}
%fffffffffffffffffffffff

The top-right panel of \autoref{fig:Ik_Ix_sdss} shows the \textit{mutual gain} as a function of length scale. The large information sharing found on small scales ($ < 10 \hmpc$) advocates the coherence of the local structures. Two points in close vicinity share a similar fate regarding the evolution of their neighbourhood; hence, there is a maximum reduction in uncertainty on smaller scales, resulting in high entropic gain. As we go towards higher length scales information sharing falls up to $60 \%$  between $10 \,-\, 20 \hmpc$. The sharp peaks that feature in the $\frac{\Delta^2 {\cal G}^{M}}{\Delta r^2}$ vs $r$ plot (bottom right panel of \autoref{fig:Ik_Ix_sdss}), correspond to the length-scale for the sharpest fall. As we can notice, this characteristic scale is slightly different for each sample and is connected to their mean intergalactic separation. Beyond this scale, the existing coherence gets rapidly weakened. In Sample 1, the average information sharing goes to zero at $\sim 135 \hmpc$. However, the mean ${\cal G}^{M}$ does not completely diminish for Samples 2 and 3 anywhere within the scope. Nevertheless, while examining the second derivatives, we observe that $\frac{\Delta^2 {\cal G}^{M}}{\Delta r^2}$ for Sample 2 and 3 gradually approaches zero, exhibiting a steadily reducing amplitude. This implies that $\frac{\Delta {\cal G}^{M}}{\Delta r}$ is getting closer to a constant value. Now, from the ever-reducing trend for the ${\cal G}^{M} (r)$, at the largest available length scales, with a constant slope, will force the mutual gain eventually diminish at a certain length scale. This scale however is beyond the explored extent for these two samples. Note that galaxies separated by a distance beyond the scale of homogeneity cannot affect the reduction in uncertainty about their prospective evolutionary trajectories through gravitational interactions, but that does not imply causal disassociation \\

\subsubsection{Information Sharing at different redshifts}
We use N-body simulations to study the information distribution at different redshifts in a $\Lambda CDM$ Universe. We went over $80$ N-body realisations of the distribution of DM particles at redshifts of $0$, $0.5$, $1$, and $2$ ( see \autoref{sec:nbody} ). For a fair comparison, we compare the results with sample 3 (SDSS), since each cube has the same size of $250 \hmpc$. \autoref{fig:Ik_Ix_nbody} shows that the overall amplitude of both information spectrum and mutual gain is much higher in sample 3 compared to the N-body cubes. One might expect the Sample 3 results to be substantially closer to $z=0.5$, but the mutual gain for Sample 3 is almost twice the $z=0.5$ distributions. The cause of this difference is the unfair comparison between galaxy and DM distributions. One has to make a fair comparison by using a mock galaxy distribution instead; this is addressed in the next section. The characteristic that stands out in this case is the information content gradually increasing across all length scales with decreasing redshifts. Additionally, the homogeneity scale for the N-body cubes drops as we go towards higher redshifts. These findings are rather typical. The offset from a homogenous random background would increase with decreasing redshift, as the structures are better defined and less arbitrary on lower redshifts. Incoherent structures with greater uncertainty are less likely to contribute to the reduction of uncertainty at other distant points in the space. Hence, the scale of homogeneity is naturally expected to drop for distributions at higher redshifts. The results in \autoref{fig:Ik_Ix_nbody} are consistent with this expectation. For all the N-body cubes homogeneity is achieved within $130 \hmpc$; whereas, the SDSS Sample 3 distribution tends towards homogeneity beyond $150 \hmpc$. An oscillating residue is noticed beyond that which may or may not originate due to the finite volume of the cube.\\

\subsubsection{Effect of linear bias and redshift-space distortions}
\autoref{fig:Ik_Ix_rsd} presents the en bloc effect of RSD and bias on the information distribution. In the left panel of the figure, we show the information spectra of the distributions with and without RSD and biasing. Let us first compare the real-space and redshift-space unbiased distributions. We observe that the redshift-space distribution exhibits suppression of information on higher frequencies, whereas enhancement in the information spectrum can be noticed for larger wavelengths. The first effect is the manifestation of the \textit{Finger-of-God} (FOG) effect, and the second one is caused by the \textit{Kaiser effect}; both of these are effects of RSD. The random peculiar motion of galaxies in dense environments such as galaxy clusters causes the elongation of the apparent shape of structures along the line of sight. This FOG effect reduces the coherence of the particles in close vicinity, that exist in the real-space distribution, causing the signal to drop on higher frequencies. On the contrary, on large scales, galaxies experience bulk infall towards large overdensities, causing galaxies to appear closer to each other than they are. Hence artificial spatial coherence is introduced in the signal on large scales. This also explains the observed fall of information for Sample 1 in the top-left panel of \autoref{fig:Ik_Ix_sdss}. Both the distance of the observer and the clustering of particles affect the degree of information suppression on large $k$. The suppression would increase with increasing density contrast and proximity of the observer relative to the cube. Unlike RSD, the effect of biasing does not affect the information in a high $k$ regime. Only the large-scale information gain is affected due to biasing. Information is found to be enhanced with biasing strength and this effect becomes more prominent as we move towards smaller $k$ values. However, the difference between the information spectrum of Sample 3 and the biased distribution for $k > 0.1 \hmpci$ is prominent and the difference increases with $k$. This indicates the necessity of non-linear and scale-dependent biasing to model the information spectrum of the observed galaxy distributions, especially for a non-linear perturbation regime.\\

\subsubsection{On the scale of homogeneity}
On the right panel of this figure, we show the information sharing for the different distributions. We also notice the enhancement in uncertainty when anti-biasing is applied, this reduces the entropic gain for the distribution with $b_1 = 0.75$, in all wavelengths. As we keep the bias increasing, the dark-matter particles are stripped away from the existing uncertainty, and a significant entropic gain is observed. For $b_1 =2$, we find the mutual gain of the redshifted distribution to come within 1-$\sigma$ of the SDSS Sample 3. Despite the differences on the largest scales, a reduced-$\chi^2$ of $0.49$ suggests that the two distributions are drawn from the same population, with a $99.8 \%$ confidence level. Both the real-space unbiased and redshift-space anti-biased distributions reach homogeneity within $120 \hmpc$ whereas the distributions with $b_1 = 1$, $b_1 = 1.5$ and $b_1 = 2$ in redshift space achieve homogeneity around $150-160 \hmpc$. Compared with the $b_1=2$ distribution one can expect the SDSS sample 3 distribution to reach homogeneity beyond $160 \hmpc$. Although large standard deviations beyond $100 \hmpc$ do not allow us to have a unanimous confirmation on a global homogeneity scale within this scope, we can at least rule out its presence in the observable Universe up to $130 \hmpc$. 

% %%%%%%%%%%%%%%%%%%%%%%%%%%%%%%%%%%%%%%%%%%%%%%%%%%%%%%%%%%%%%%%%%%%%%%%%%%%%%%%%%%%%%%%%%%%%%%%%%%%%%%%%%%%%%%%%%%

% %%%%%%%%%%%%%%%%%%%%%%%%%%%%%%%%%%%%%%%%%%%%%%%%%%%%%%%%%%%%%%%%%%%%%%%%%%%%%%%%%%%%%%%%%%%%%%%%%%%%%%%%%%%%%%%%%%

\section{Conclusion}
%%%%%%%%%%%%%-----------------------------------------------------%%%%%%%%%%%%%%%%
The two-point correlation function and its Fourier counterpart the power spectrum are the widely used statistics for studying galaxy clustering. However, these two estimators and their multiple variants that rely upon a finite number of moments of density perturbation, may not be sufficient in analysing the galaxy and dark-matter clustering in the non-linear regime. We introduce the \textit{entropic gain}, an information-theoretic identity that uses all the higher-order moments of perturbation in analysing the nature of clustering in the observable galaxy distribution. We employ this tool to analyse the real and Fourier space distribution of information in 3 different galaxy samples prepared using data from SDSS, each representing a specific redshift zone and a particular class of galaxy. We also quantify the information reduction mutually caused by structures at any two points in space, in the accessible volume. We search for a universal scale of homogeneity, where ( by definition ) the mutual gain in information diminishes irrespective of the local clustering strength. We don't find such a characteristic scale at least up to $130 \hmpc$. Furthermore, to study the evolution of the Universe's information content, we study a log-normal density distribution that evolves with time. Redshift-dependent variance, skewness and kurtosis mimic the probability distribution of the actual density contrast in the matter density field. It is found that the growth of structures inevitably causes a reduction in information entropy. Entropy being one of the fundamental physical entities in the Universe could cause the observed accelerated expansion itself. For a thermodynamic system, the increment in the volume allows it to have a larger number of microstates to access. A configuration with a higher volume is preferred by a system due to its innate inclination towards achieving a higher entropy state. The nonlinear reduction in entropy due to structure formation can indirectly cause the accelerated expansion of the space. However, here we are measuring merely the drop in information entropy. So the theory of the accelerated expansion being caused by a growing entropic drop due to structure formation requires a deeper inspection. Although we have limited our discussion to the $\Lambda CDM$ cosmology, it is possible to study the entropic gain of different cosmological models and constrain the cosmological parameters with the aid of appropriate observables. This paper is meant to be the initial step, laying the foundation for future research that will build on it and will be better equipped to address these issues. 

% %%%%%%%%%%%%%%%%%%%%%%%%%%%%%%%%%%%%%%%%%%%%%%%%%%%%%%%%%%%%%%%%%%%%%%%%%%%%%%%%%%%%%%%%%%%%%%%%%%%%%%%%%%%%%%%%%%
\section{Data-availability}
%%%%%%%%%%%%%-----------------------------------------------------%%%%%%%%%%%%%%%%
The data utilized in this study can be found in the SDSS database, which is open for public access. Data generated in this work can be shared on request to the author.

% %%%%%%%%%%%%%%%%%%%%%%%%%%%%%%%%%%%%%%%%%%%%%%%%%%%%%%%%%%%%%%%%%%%%%%%%%%%%%%%%%%%%%%%%%%%%%%%%%%%%%%%%%%%%%%%%%%
\section{Acknowledgement}
%%%%%%%%%%%%%-----------------------------------------------------%%%%%%%%%%%%%%%%
I thank Prof. Somnath Bharadwaj, Prof. Sayan Kar, Dr. Biswajit Pandey, Dr. Maret Einasto and fellow researchers at IIT-Kharagpur, for their suggestions and useful discussions. Special thanks to DST-SERB for support through the National Post-Doctoral Fellowship (PDF/2022/000149). Funding for the Sloan Digital Sky Survey IV has been provided by the Alfred P. Sloan Foundation, the U.S. Department of Energy Office of Science, and the Participating Institutions.  SDSS-IV acknowledges support and resources from the Center for High Performance Computing at the University of Utah. The SDSS website is www.sdss.org. SDSS-IV is managed by the Astrophysical Research Consortium for the Participating Institutions of the SDSS Collaboration including the Brazilian Participation Group, the Carnegie Institution for Science, Carnegie Mellon University, Center for Astrophysics | Harvard \& Smithsonian, the Chilean Participation Group, the French Participation Group, Instituto de Astrof\'isica de Canarias, The Johns Hopkins University, Kavli Institute for the Physics and Mathematics of the Universe (IPMU) / University of Tokyo, the Korean Participation Group, Lawrence Berkeley National Laboratory, Leibniz Institut f\"ur Astrophysik Potsdam (AIP),  Max-Planck-Institut f\"ur Astronomie (MPIA Heidelberg), Max-Planck-Institut f\"ur Astrophysik (MPA Garching), Max-Planck-Institut f\"ur Extraterrestrische Physik (MPE), National Astronomical Observatories of China, New Mexico State University, New York University, University of Notre Dame, Observat\'ario Nacional / MCTI, The Ohio State University, Pennsylvania State University, Shanghai Astronomical Observatory, United Kingrand-designom Participation Group, Universidad Nacional Aut\'onoma de M\'exico, University of Arizona, University of Colorado Boulder, University of Oxford, University of Portsmouth, University of Utah, University of Virginia, University of Washington, University of Wisconsin, Vanderbilt University, and Yale University. 

%%%%%%%%%%%%%%%%%%%%%%%%%%%%%%%%%%%%%%%%%%%%%%%%%%%%%%%%%%%%%%%%%%%%%%%%%%%%%%%%%%%%%%%%%%%%%%%%%%%%%%%%%%%%%%%%%%%%%
\bibliographystyle{JHEP}
\bibliography{info_gain.bib}
\appendix 

%%%%%%%%%%%%%%%%%%%%%%%%%%%%%%%%%%%%%%%%%%%%%%%%%%%%%%%%%%%%%%%%%%%%%%%%%%%%%%%%%%%%%%%%%%%%%%%%%%%%%%%%%%%%%%%%%%%%%
\section{Appendix:}
\label{sec:appendix}

\subsection[Derivatives of EWMs]{Finding $n^{th}$ order derivative of $k^{th}$ EWM}
\label{sec:appendix_1}
Let us start by writing the $k^{th}$ EWM as  
\begin{eqnarray}
\label{eq:A1}
{\cal A}_k = \frac{U_k}{U_0} 
\end{eqnarray}
Where 
\begin{eqnarray}
\label{eq:A2}
U_k = \int_{\mathbb{R}^3} \eta^k e^{-\eta} d^3 x 
\end{eqnarray}
\noindent By definition $\eta(\mathbf{x}, a) = D_{+} (a) \frac{\delta(\mathbf{x})}{\delta_c}$; hence by denoting  $\,\,\frac{d^q}{da^q}$  as  $\mathbb{D}^q\,\,$  and  $\,\,\frac{1}{D_{+}} \frac{\partial^q D_{+} }{\partial a^q}$  as  $Q_q\,\,$ we can write
\begin{eqnarray}
\label{eq:A3}
\mathbb{D}^q \eta =  \eta Q_q.
\end{eqnarray}
 Note that $D_{+}$ being a function of $a$ alone allows the partial and total derivatives to have interchangeable connotations. \autoref{eq:A3} also leads to the identities
\begin{eqnarray}
Q_0 &=& 1, \nonumber\\
Q_1 &=& \frac{f(\Omega_m)}{a}, \nonumber\\
Q_{n+1} &=& \mathbb{D}^1 Q_n +  Q_n Q_1, \nonumber
\end{eqnarray}

\noindent Using \autoref{eq:A2} and \autoref{eq:A3} the first derivative of $U_k$ is found as 
\begin{eqnarray}
\label{eq:A4}
\mathbb{D}^1 U_k = Q_1\, [\, k U_k - U_{k+1}\,]. 
\end{eqnarray}
The $1^{st}$ derivative of ${\cal A}_k$ then would be
\begin{eqnarray}
\label{eq:A5}
\mathbb{D}^1 {\cal A}_k &=&  \left[\, \frac{\mathbb{D}^1 U_k}{U_0} - \frac{U_k \,\mathbb{D}^1 U_0}{U^2_0} \,\right]  \nonumber \\
&=& Q_1\, [\, k {\cal A}_k - {\cal A}_{k+1} + {\cal A}_1 {\cal A}_k\,] 
\end{eqnarray}

\noindent Any the $n^{th}$ order derivative of the $k^{th}$ EWM can now be estimated from the general recurrence relation
\begin{eqnarray}
\label{eq:EWA_diff}
\mathbb{D}^n {\cal A}_k &=& \sum_{q=0}^{n-1} \,\, \Comb{n-1}{q}\,\,\, \mathbb{F}^{n-q-1}_{1} \cdot \mathbb{T}^q_k.
\end{eqnarray}
\begin{eqnarray}
\text{where} \,\,\, \mathbb{F}^{n-q-1}_{s} &=& \mathbb{D}^{n-q-1} Q_s\,\, \nonumber \\
\text{and} \,\,\, \mathbb{T}^q_k &=& k \, \mathbb{D}^q{\cal A}_k\,- \,\mathbb{D}^q{\cal A}_{k+1} \,+\, \sum_{j=0}^{q} \Comb{q}{j}\,\, \mathbb{D}^{q-j}{\cal A}_{k} \,\,\mathbb{D}^j {\cal A}_{1} \nonumber
\end{eqnarray}

\noindent One can find the following identities using \autoref{eq:EWA_diff}. 
\begin{eqnarray}
\label{eq:Deriv_12}
\mathbb{D}^1 {\cal A}_1 & = & Q_1 \left\{ {\cal A}_1 - {\cal A}_2 + {\cal A}^2_1\right\} \\
\mathbb{D}^1 {\cal A}_2 & = & Q_1 \left\{ 2 {\cal A}_2 - {\cal A}_3 + {\cal A}_2 {\cal A}_1 \right\}  \\
\mathbb{D}^2 {\cal A}_1 & = &  Q_1^2 \left\{ 2{\cal A}_1^3 + {\cal A}_1^2 - 3{\cal A}_1 {\cal A} _2 - 2{\cal A}_2 + {\cal A}_3 \right\} + Q_2 \left\{ {\cal A}_1 - {\cal A}_2 + {\cal A}^2_1\right\}  
\end{eqnarray}

%#-------------------------------------------------------------------------------------------#
\subsection[Shot noise correction]{Shot noise correction for entropic gain}
\label{sec:appendix_2}
Consider a discrete distribution of ${\cal N}$ galaxies confined in a finite cubic volume, where each of the voxels accommodates $\bar{f}_{\cal M}\,\cal{N}$ galaxies on average. From \autoref{eq:sp_info_gain} for any $i^{th}$ voxel we get the specific information gain as
\begin{eqnarray}
\label{eq:sp_info_gain2}
\Phi \,(\,\mathbf{x}_i\, ) &=&  \frac{1}{\bar{I}\,U_0\,\Delta x^3} \biggl[\,\,\, \Delta I\,(\delta\, ( \mathbf{x}_i)\, ) \,\, \exp \{\,- \Delta I\,(\,\delta (\mathbf{x}_i)\,) \,\} \,\,\,\biggr].
\end{eqnarray}
where $\Delta I^{(i)}= \frac{\delta (\mathbf{x}_i)}{\delta_c}$.\\ 

\noindent With $n = (1+\delta)\,\bar{f}_{\cal M}\,\cal{N}$ as the count in a given voxel and $\sigma^2_n$ as the variance of the galaxy count, we estimate the shot-noise corrected $\Delta I$, i.e.
\begin{eqnarray}
\label{eq:DI_SNC1}
\Delta I_{SNC} (\delta) = \Delta I (\delta) - \left| \frac{\partial \, \Delta I}{\partial n} \right| \sigma_{n}
\end{eqnarray}
We also have  $\frac{\partial \, \Delta I}{\partial n} = \frac{1}{\delta_c\,\bar{f}_{\cal M}\,\cal{N}}$ and $\sigma_n = \sqrt{\bar{f}_{\cal M}\,\cal{N}}$  assumimng the particle count follows a Poisson distribution. Hence, the shot noise correction would give us
\begin{eqnarray}
\label{eq:G_SNC}
\Phi^{(i)}_{SNC} (\eta) &=& \frac{1}{\bar{I} \, U_0^{SNC} \Delta x^3} \left[ \Delta I^{(i)}_{SNC}\, \exp\{\, -\Delta I^{(i)}_{SNC}\,\} \right]
\end{eqnarray}
with $U_0^{SNC} = \sum_{i} \exp \{ - \Delta I_{SNC}^{(i)} \}$ and 
\begin{eqnarray}
\label{eq:DI_SNC2}
\Delta I^{(i)}_{SNC} = \Delta I^{(i)} - \frac{1}{\delta_c \sqrt{\bar{f}_{\cal M}\,\cal{N}}}  
\end{eqnarray}
%#-------------------------------------------------------------------------------------------#

\subsection[Scale independent bias estimation]{Scale independent estimation of linear bias}
\label{sec:appendix_3}

Let us consider the normalised probability distributions of $\delta$ for the unbiased and biased distributions to be $p^u(\delta)$ and $p^b(\delta)$ respectively. We divide the entire range of $\delta$ into $N_\delta$ bins. If the probabilities for the unbiased distribution in the $i^{th}$ and $j^{th}$ bins of $\delta$ are $p_i^u$ and $p_j^u$ respectively, then the joint probability of $(\delta_i,\delta_j)$ pairs to be found in the distribution would be proportional to $p^u_i \delta^i \cdot p^u_j \delta^j$. Using these joint probabilities, we propose a scale-independent linear bias estimator 
\begin{eqnarray}
b_1 = \sqrt{\frac{\sum_i \sum_j p^u_i \delta_i \cdot p^u_j \delta_j}{\sum_i \sum_j p^b_i \delta_i \cdot p^b_j \delta_j}}
\end{eqnarray}
to measure the average linear bias of the newly formed distribution obtained after applying the selection function. Note that, $p^u$ and $p^b$ both are normalized so $\sum_i \sum_j p^u_i p^u_j = \sum_i \sum_j p^b_i p^b_j =1$. We choose $N_{\delta} = 1000$ for this work. 

\begin{center}
\rule{0.2\textwidth}{1pt}
\end{center}
\label{lastpage}
\end{document}